\documentclass[preprint,12pt, nofootinbib]{revtex4}
\usepackage[brazil, english]{babel}
\usepackage[utf8]{inputenc}
\usepackage{ulem}
\usepackage{graphicx}
\usepackage{epsfig}
\usepackage{amssymb}
\usepackage{amsmath} 
\usepackage{slashed}
\usepackage[unicode=true,bookmarks=false,breaklinks=false,pdfborder={0 0 1},colorlinks=true]
 {hyperref}
\hypersetup{
 citecolor=blue,linkcolor=blue,urlcolor=blue}

\graphicspath{%
    {converted_graphics/}
    {/}
}
\begin{document}

\title{Magnetized AdS/BCFT Correspondence in Horndeski Gravity} 

\author{Fabiano F. Santos$^{1}$}
\email[Eletronic address: ]{fabiano.ffs23@gmail.com}
\author{Mois\'es Bravo-Gaete$^{2}$}
\email[Eletronic address: ]{mbravo@ucm.cl} 
\author{Manoel M. Ferreira$^{1}$}
\email[Eletronic address: ]{manojr.ufma@gmail.com}
\author{Rodolfo Casana$^{1}$}
\email[Eletronic address: ]{rodolfo.casana@gmail.com}
\affiliation{$^1$Departamento de Física, Universidade Federal do Maranhão, Campus Universitario do Bacanga, São Luís (MA), 65080-805, Brazil.\\
$^2$Departamento de Matem\'atica, F\'isica y Estad\'istica, Facultad de Ciencias
B\'asicas, Universidad Cat\'olica del Maule, Casilla 617, Talca, Chile.}

\begin{abstract}
This work examines the thermodynamics and hydrodynamics behaviors of a five-dimensional black hole under the influence of an external magnetic field. The solution is the gravity dual to the Anti-de Sitter/Boundary Conformal Field Theory correspondence, enabling the study of properties within an anisotropic fluid framework. Utilizing holographic renormalization, we compute the free energy and the holographic stress tensor residing on the boundary denoted as $Q$. Within the fluid/gravity correspondence framework, we have a class of boundary extensions in $Q$, where the stress-energy tensor describes a magnetizing conformal fluid. We discuss the characteristics of this special solution as well as its thermodynamic properties, including the bulk and shear viscosity, the square of the speed of sound, as well as the anisotropic effects induced by the magnetic field in the magnetized conformal plasma.
\end{abstract}

\maketitle

\section{Introduction}
In recent years, the description of macroscopic properties of strongly coupled matter has been a significant challenge, requiring the use of non-perturbative methods and related via gravity, thanks to the Anti-de Sitter/Conformal Field Theory (AdS/CFT) correspondence \cite{Maldacena:1997re,Witten:1998qj}. This correspondence maps a $D-$dimensional strongly coupled fluid which enjoys conformal symmetry at finite temperature to a $(D+1)$-asymptotically AdS black hole (BH). 

An illustrative example lies in the study of matter in the plasma state, more specifically Quark-Gluon Plasma (QGP), produced in collisions of heavy ions at the Relativistic Heavy Ion Collider and Large Hadron Collider. Here, investigating the thermodynamic and hydrodynamic properties of Quantum Chromodynamics (QCD) at high temperatures is of paramount importance, given the relevance of non-perturbative effects  (see, for example Refs. \cite{Ahn:2022azl,Ballon-Bayona:2022uyy,Skokov:2009qp,Fukushima:2016xgg,Gubser:2008px}). For the four-dimensional situation, the $\mathcal{N}=4$ Super Yang-Mills theory \cite{Gubser:1996de,Burgess:1999vb,Policastro:2001yc,Policastro:2002se} has become an excellent laboratory to explore macroscopic properties at finite temperatures, such as the shear viscosity to entropy density ratio \cite{Policastro:2001yc,Policastro:2002se}, which matches closely to the expected results for the QGP observed in heavy ion collisions.
\begin{figure}[!htbp]
    \centering
    \includegraphics[scale=0.5]{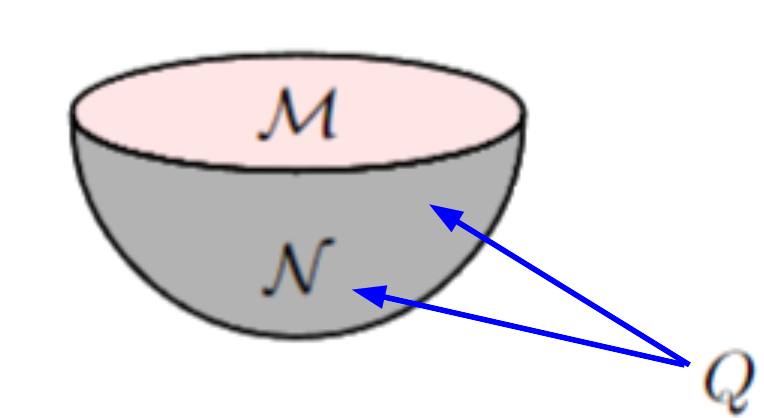}\quad \quad
    \includegraphics[scale=0.28]{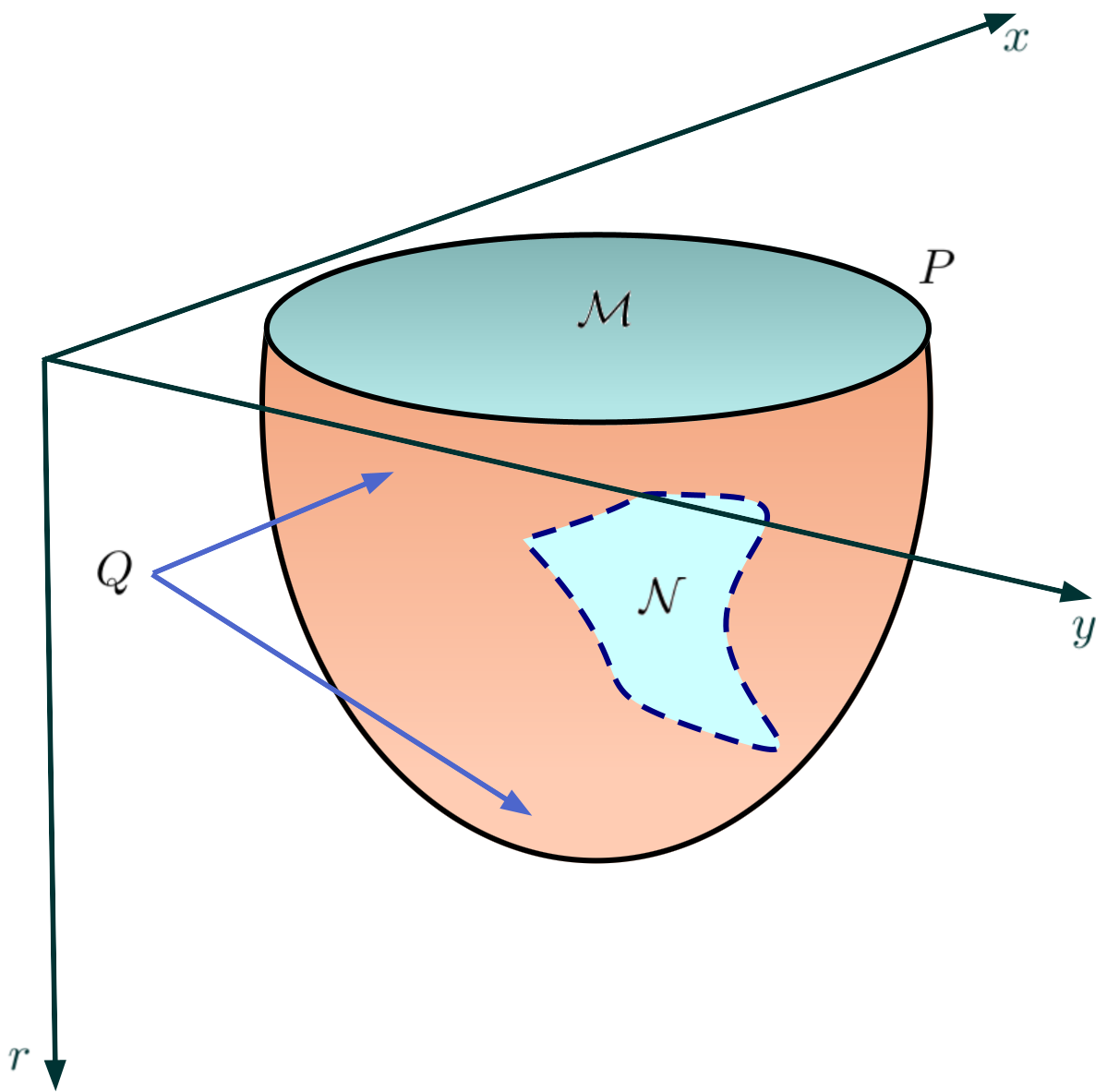}
    \caption{{\bf{Left Panel:}} Graphic representation of the AdS/BCFT correspondence illustrates the relationship between the components involved. Here, $\mathcal{M}$ denotes the manifold hosting the CFT, with its boundary defined as $\partial \mathcal{M}$. The gravity dual, denoted by $\mathcal{N}$, is such that its asymptotic boundary corresponds to $\mathcal{M}$. Together with the above, $\partial \mathcal{M}$ is extended into the bulk AdS, which constitutes the boundary of the $D-$dimensional manifold $Q$. {\bf{Right Panel:}} For this graphic representation, $\mathcal{N}$ is the subspace of the bulk of AdS$_{D+1}$, bounded by $Q$ which it encodes physics of $\mathcal{M}$. $P$ is the common boundary of $Q$ and $\mathcal{M}$.}
    \label{B}
\end{figure}

Building on this idea, there has been a rise in interest in expanding the AdS/CFT duality in recent years, leading to the development of important extensions such as the holographic duality known as Anti-de Sitter/Boundary Conformal Field Theory (AdS/BCFT) correspondence \cite{Takayanagi:2011zk,Fujita:2011fp,Fujita:2012fp}. This extension introduces a novel scenario where the CFT is defined on a manifold $\mathcal{M}$ with a boundary $\partial \mathcal{M}$. Therefore, in the holographic dual, the manifold boundary of a $D$-dimensional manifold $\mathcal{M}$ corresponds to a $(D+1)$-dimensional asymptotically AdS space $\mathcal{N}$ with $\partial \mathcal{N}=\mathcal{M} \cup Q$. Here, $Q$ corresponds to a $D$-dimensional manifold that satisfies $\partial Q=\partial \mathcal{M}$ (see Figure \ref{B}, left panel).

To explore the AdS/CFT correspondence, we need to impose the Dirichlet boundary condition (DBC) at the boundary of AdS, for then perform the DBC on $\mathcal{M}$. But, according to \cite{Takayanagi:2011zk,Fujita:2011fp}, for the AdS/BCFT duality, a Neumann boundary condition (NBC) on $Q$ is required and, from the standpoint of holography, this boundary should be dynamic \cite{Fujita:2011fp}. Such dynamics can be introduced through the specification of the boundary conditions of the variational problem. In recent years, this framework has garnered attention for its novel approach to computing transport coefficients, where BHs play a crucial role. Examples include the Hawking-Page phase transition, the Hall conductivity, fluid/gravity correspondence in Einstein gravity \cite{Fujita:2011fp,Melnikov:2012tb,Miao:2018qkc,Magan:2014dwa} and its extensions \cite{dosSantos:2022scy,Sokoliuk:2022llp,Santos:2021orr,Santos:2023flb}. In addition to the aforementioned, the nature of the AdS/BCFT duality is deeply ingrained in the holographic computation of entanglement entropy of Hawking radiation in dyonic BHs \cite{Jeong:2023hrb}, within the frameworks of Einstein gravity \cite{Ryu:2006bv}, Horndeski gravity \cite{DosSantos:2022exb,Santos:2023eqp}, and the Randall-Sundrum (RS) model \cite{Randall:1999vf}.

This extension of the CFT’s boundary inside the bulk of the AdS-space is a modification of a \textit{thin} RS  brane, which intersects the AdS boundary. This concept is applicable in theories such as Horndeski models \cite{Santos:2021orr,Santos:2023flb}. In this scenario, the RS brane is a dynamic object, where a NBC discontinuity in the bulk extrinsic curvature across the defect is compensated by the tension from the brane. These boundaries are known as the RS branes. In the right panel of Figure \ref{B}, we illustrate the boundary denoted as $P$ and determined by the condition $y = \text{constant}$. This boundary represents one of the coordinates on $\mathcal{M}$, corresponding to the AdS/BCFT problem, considering over half of Minkowski space. The solution with $y = \text{constant}$ predicts the presence of gravity solutions with non-zero tension for the RS branes \cite{Santos:2021orr,Santos:2023flb}. Recent studies have demonstrated the existence of such solutions, exploring their potential to describe charged BHs \cite{Santos:2023flb}.

By employing the AdS/BCFT framework, we can break conformal symmetry and introduce a single scalar field from Horndeski gravity into the system \cite{Horndeski:1974wa,Deffayet:2009mn,Charmousis:2014mia,Rinaldi:2012vy}. These scalar fields possess non-zero profiles in the bulk, allowing us to study the transport of coefficients such as bulk and shear viscosity, denoted as $\zeta$ and $\eta$ respectively (see for example Refs. \cite{Santos:2023flb,Feng:2015oea,Bravo-Gaete:2022lno,Bravo-Gaete:2023iry,Bravo-Gaete:2020lzs,Parnachev:2005hh,Benincasa:2005iv,Buchel:2005cv,Benincasa:2006ei,Mas:2007ng,Bravo-Gaete:2021hlc}). For these works, the ratios $\zeta/S$ and $\eta/S$, where
the first is responsible for the equilibrium of a fluid subject to small expansion or compression, and the second is associated with energy dissipation due to the relative movement of the fluid layers \cite{Torres-Rincon:2012sda,Mishra:2020onx}, are affected via the contributions of Horndeski gravity, magnetic field, temperature, and the profile $Q$ through of $S$,  where $S$ represents the density entropy. The advantage of these procedures is that results are analytic, and in agreement with the numerical results obtained previously in \cite{Gubser:2008sz}. Specifically, at higher temperatures, the fluid transitions to the plasma phase where $\zeta/S\to0$ and $\eta/S\to0$, closely resembling the proposition in \cite{Gubser:2008sz}. Furthermore, in the absence of a magnetic field, denoted as $B$, these ratios are violated, as anticipated for a strongly coupled anisotropic plasma \cite{Rebhan:2011vd}, characterized by anisotropic pressures \cite{Ahn:2022azl,Ballon-Bayona:2022uyy}. 

The square of the speed of sound, denoted by $c_s^2$, is a crucial parameter influencing the acceleration of a fluid \cite{Bluhm:2007nu}. In a category of four-dimensional field theories that are strongly coupled, $c_s^2$ has been bounded above by a value of $1/3$ at high temperatures (known as the conformal bound), as shown in \cite{Cherman:2009tw}. Nevertheless, the above bound is affected when additional conditions are considered, such as at low temperature and high density \cite{Iida:2022hyy,Itou:2023pcl} , as well as with non-zero isospin chemical potential \cite{Brandt:2022hwy,Abbott:2023coj}.

In light of everything mentioned above, our study leverages the $\text{AdS}_5/\text{BCFT}_4$ correspondence to delve into the thermodynamic and hydrodynamic characteristics of strongly coupled $\mathcal{N}=4$ Super Yang-Mills plasma under the influence of a magnetic field. For this, we consider a five-dimensional scenario within Horndeski gravity, coupled with an external magnetic field $B$, capturing the essential degrees of freedom needed to match an equation of state with a range of BHs solutions.

To achieve this, the free energy and the thermodynamic quantities play a providential role, in particular for the entropy, allowing us to compute the transport coefficients $c_s^2$, $\zeta/S$, $\eta/S$ and examining their responses to gravitational theory and the presence of $B$ \cite{Buchel:2007mf,Gubser:2008yx,Kharzeev:2007wb,Karsch:2007jc,Kharzeev:2007jp,Fukushima:2008xe,Endrodi:2018ikq}. Via holographic renormalization, we describe a family of boundary stress-energy tensors residing in $Q$, consistent with the asymptotically AdS$_{5}$ BH in the bulk. Each of these tensors corresponds to a hypersurface in the volume that bounds a subspace of the BH solution, allowing us to extract the hydrodynamic transport coefficients such as energy density, and longitudinal and transverse pressure.

Our analysis reveals that the observed anisotropy, due to the presence of the magnetic field, is consistent with expectations derived from experimental observations of QGP \cite{Skokov:2009qp}. The anisotropic hydrodynamic effects lead us to the fluid/gravity correspondence within the AdS$_{5}$/BCFT$_{4}$ scenario.

This work is organized as follows: In Section \ref{v1}, we consider the gravitational setup, showing the solution and the $Q$-boundary profile. In Section \ref{sec:HR}, we compute the Euclidean on-shell action, which is related to the free energy of the corresponding thermodynamic system, where in particular, we will focus on the entropy, to then in Section \ref{sec:hol} to obtain the bulk viscosity, shear viscosity, and the speed of sound. In section \ref{fluid/grav}, we present the fluid/gravity correspondence. Finally, Section \ref{v4} is devoted to our conclusions and discussions.

\section{The setup, equations of motion and the $Q$-boundary profile}\label{v1}

To study the transport coefficients in the presence of a magnetic field, with this configuration, the action containing all the necessary components for our description is given by
\begin{eqnarray}\label{açao}
S_{bulk}&=&S^{\mathcal{N}}_{\rm H}+S^{\mathcal{N}}_{\rm M}+S^{\mathcal{N}}_{\rm 2-FF}+S^{\mathcal{N}}_{mat},\nonumber\\
&=&\int_{\mathcal{N}}d^{5} x \sqrt{-g}\; \left(\kappa {\cal L}_{\rm H}+\kappa {\cal L}_{\rm M}+\lambda^2 {\cal L}_{\rm 2-FF}+{\cal L}_{mat}\right),\label{1}
\end{eqnarray}
where $\kappa={1}/{(16 \pi G_N)}$, with $G_N$ is the Newton Gravitational constant, $\lambda^2$ a coupling constant, and
\begin{eqnarray}
{\cal L}_{\rm H}&=&   (R-2\Lambda)\label{L1} -\frac{1}{2}(\alpha g_{\mu\nu}-\gamma\,  G_{\mu\nu})\nabla^{\mu}\phi\nabla^{\nu}\phi ,\label{eq:Lhorn}\\
{\cal L}_{\rm M}&=&-\dfrac{1}{4e^{2}} F^{\mu \nu} F_{\mu \nu}, \label{eq:Maxwell}\\
{\cal L}_{\rm 2-FF}&=&-\frac{1}{12} (dM)^2 - \frac{m^2}{4}M^{\mu \nu} M_{\mu \nu}- \frac{1}{2}M^{\mu \nu}F_{\mu \nu}-\frac{J}{8}V(M).\label{eq:L2FF}
\end{eqnarray}
Here, for ${\cal L}_{\rm H}$, we have that $R=g^{\mu \nu} R_{\mu \nu}$, $G_{\mu \nu}$ and $\Lambda$ represent the scalar curvature, the Einstein tensor, and the cosmological constant respectively, while that $\phi=\phi(r)$ is a scalar field, $\alpha$, and $\gamma$ are coupling constants. It is interesting to note that Lagrangian (\ref{eq:Lhorn}) has been explored from the point of view of hairy black hole configurations 
\cite{Rinaldi:2012vy,Babichev:2013cya,Anabalon:2013oea,Bravo-Gaete:2014haa,Bravo-Gaete:2013dca}, boson and neutron stars \cite{Brihaye:2016lin,Cisterna:2015yla,Cisterna:2016vdx}, Hairy Taub-NUT/Bolt-AdS solutions \cite{Arratia:2020hoy}, holographic renormalization \cite{Caceres:2023gfa},  as well as holographic applications such that quantum complexity and shear viscosity 
\cite{Feng:2018sqm,Feng:2015oea,Bravo-Gaete:2022lno,Bravo-Gaete:2021hlc,Bravo-Gaete:2020lzs}. ${\cal L}_{\rm M}$ represents the Maxwell Lagrangian, where $F_{\mu \nu}=\partial_{\mu} A_{\nu}-\partial_{\nu} A_{\mu}$ and $e$ is a coupling constant. The Lagrangian ${\cal L}_{\rm 2-FF}$ is constructed through a 2-form $M_{\mu\nu}$, where $dM=(dM)_{\tau \mu \nu} =3 \nabla_{[\tau} M_{\mu \nu]}$ is the exterior differential and $(dM)^2=9 \nabla_{[\tau} M_{\mu \nu]} \nabla^{[\tau} M^{\mu \nu]}$. $V(M)$ describes the self-interaction of polarization tensor, with $J$ a constant, and $m$ is a constant related to the mass. Finally, $S^{\mathcal{N}}_{mat}$ is the action associated with matter sources.

Under this scenario, to establish the $\mbox{AdS}_{5}/\mbox{BCFT}_{4}$ correspondence, we need to construct the terms of the boundary. Following the Refs. \cite{Santos:2021orr,Santos:2023flb}, these expressions are given by 
\begin{eqnarray}
S_{BCFT}&=&{2\kappa\int_{Q}{d^{4}x\sqrt{-h}\mathcal{L}_{bdry}}+2\int_{Q}{d^{4}x\sqrt{-h}\mathcal{L}_{mat}}+2\kappa\int_{ct}{d^{4}x\sqrt{-h}\mathcal{L}_{ct}}}\nonumber\\
&+&S^{Q}_{mat},\label{eq:BCFT}
\end{eqnarray}
with
\begin{eqnarray}
\mathcal{L}_{bdry}&=&(K-\Sigma)-\frac{\gamma}{4}(\nabla_{\mu}\phi\nabla_{\nu}\phi n^{\mu}n^{\nu}-(\nabla \phi)^2)K-\frac{\gamma}{4}\nabla_{\mu}\phi\nabla_{\nu}\phi K^{\mu\nu},\label{eq:Lbound}\\
\mathcal{L}_{ct}&=&c_{0}+c_{1}R+c_{2}R^{ij}R_{ij}+c_{3}R^{2}+b_{1}(\partial_{i}\phi\partial^{i}\phi)^{2}+\cdots.
\end{eqnarray}
For the Lagrangian $\mathcal{L}_{bdry}$, $K_{\mu\nu}=h^{\phantom{\mu}\beta}_{\mu}\nabla_{\beta}n_{\nu}$ corresponds to the extrinsic curvature where $K=h^{\mu\nu}K_{\mu\nu}$ is the trace, $h_{\mu\nu}$ is the induced metric while that $n^{\mu}$ is an outward pointing unit normal vector to the boundary of the hypersurface $Q$. Additionally, $\Sigma$ is the boundary tension on $Q$ and $S^{Q}_{mat}$ is the matter action on $Q$. ${\cal L}_{ct}$ represents the boundary counterterms, which do not influence the bulk dynamics and hence will be disregarded.

With this previous presentation from the bulk and boundary side for the AdS/BCFT correspondence, from eqs. (\ref{1}) and (\ref{eq:BCFT}) we can present the total action $S$ as
\begin{eqnarray}\label{açao}
S &=& S_{bulk}+S_{BCFT},\label{eq:Stot}
\end{eqnarray}
and with respect to the equations of motions for the action (\ref{eq:Stot}), they can be delineated into three crucial segments: (i) the Einstein-Horndeski equations, (ii) the equations to provide the profile solution (see for example, Refs. \cite{Magan:2014dwa,Melnikov:2012tb,Fujita:2011fp,Takayanagi:2011zk,Santos:2021orr}), and (iii) the equations of motion for the electromagnetic sector. This last part is provided through probe approximation, as discussed in \cite{Santos:2023flb}.

As a first step, we start imposing the NBC where according to \cite{Santos:2021orr,Santos:2023flb} takes the form
\begin{eqnarray}
K_{\alpha\beta}-h_{\alpha\beta}(K-\Sigma)-\frac{\gamma}{4}H_{\alpha\beta}=\kappa {\cal S}^{Q}_{\alpha\beta}\,,\label{L7}
\end{eqnarray}
with 
\begin{eqnarray}
&&H_{\alpha\beta}:=(\nabla_{\sigma}\phi\nabla_{\rho}\phi\, n^{\sigma}n^{\rho}-(\nabla \phi)^2) (K_{\alpha\beta}-h_{\alpha\beta}K)-(\nabla_{\alpha}\phi\nabla_{\beta}\phi)K\,,\label{L8}
\end{eqnarray}
and ${\cal S}^{Q}_{\alpha\beta}$ represents the variation of the action $S^{Q}_{mat}$ with respect to the induced metric $h_{\mu \nu}$, this is, ${\cal S}^{Q}_{\alpha\beta}=-({2}/{\sqrt{-h}})({\delta S^{Q}_{mat}}/{\delta h^{\alpha\beta}})$. Here, we consider the matter stress-energy tensor on $Q$ as a constant, implying that ${\cal S}^{Q}_{\alpha\beta}=0$.

On the other hand, from the Einstein-Horndeski model (\ref{1})-(\ref{eq:Lhorn}), assuming that $S^{\mathcal{N}}_{mat}$ is constant, the equations of motions on $S^{\mathcal{N}}_{\rm H}$ and $S^{Q}_{bdry}$ with respect to the dynamical fields $g_{\mu \nu}$ and $\phi$ are given by ${\cal E}_{\mu\nu}=0$, $  {\cal E}_{\phi}=0$ and ${\cal F}_{\phi}=0$, where the explicit expressions are reported in Appendix \ref{AA}.

In this model, we focus on a static BH. The approach outlined in Refs. \cite{Rinaldi:2012vy,Babichev:2013cya,Anabalon:2013oea,Bravo-Gaete:2014haa,Bravo-Gaete:2013dca} enables us to derive static BH configurations, thus bypassing the no-hair theorems \cite{Hui:2012qt}. For this particular scenario, it is essential that $J^{r}_{\phi}=0$, while still allowing flexibility in the radial dependence of the scalar field $\phi$. This condition can be expressed using eq. (\ref{L11}):
\begin{equation}
J^{r}_{\phi}=\alpha g_{rr}-\gamma G_{rr}=0\label{L13}\,, 
\end{equation}
and defining $\phi{'}(r):= {\psi}(r)$, where $ (')$ denotes the derivative with respect to the radial coordinate $r$, we can first show that the equation ${\cal E}_{\phi}=0$ is trivially satisfied. For this setup, considering the five-dimensional metric 
\begin{equation}\label{ansatz}
ds^2= \frac{L^2}{r^2}\left(-f(r)\,dt^2+dx^2+dy^2+dw^2+\frac{dr^2}{f(r)}\right),
\end{equation}
where $ x_1 \leq x \leq x_2$, $ y_1 \leq y \leq y_2$ and $ w_1 \leq w \leq w_2$, the metric function $f(r)$ from eq. (\ref{L13}) takes the form \cite{Santos:2021orr,Bravo-Gaete:2014haa} 
\begin{eqnarray}
&&f(r)=\frac{\alpha L^{2}}{3\gamma}\left[1-\left(\frac{r}{r_{h}}\right)^{4}\right].\label{L14}
\end{eqnarray}
Here, the integration constant $r_h$ represents the location of the event horizon, while the remainder equations of motions are satisfied when $\psi(r)$ reads
\begin{eqnarray}
&&\psi^{2}(r)=(\phi'(r))^2=-\frac{2L^{2} \xi }{\gamma r^{2}f(r)}\,,\label{L15}
\end{eqnarray}
where we define
\begin{equation}\label{eq:xi}
\xi=\dfrac{\alpha+\gamma\Lambda}{\alpha},    
\end{equation}
and the scalar field is real, only if
$$\alpha+\Lambda \gamma \leq 0.$$
For the sake of completeness, following the steps of Refs. \cite{Santos:2021orr,Brito:2019ose}, via the transformations
\begin{eqnarray}
&&f(r) \rightarrow \frac{\alpha L^{2}}{3\gamma} f(r),\qquad t \rightarrow \frac{3\gamma}{\alpha L^{2}} t,\qquad\,w\rightarrow \sqrt{\frac{3\gamma}{\alpha L^{2}}} w,\nonumber\\
&&x \rightarrow \sqrt{\frac{3\gamma}{\alpha L^{2}}} x, \qquad y \rightarrow \sqrt{\frac{3\gamma}{\alpha L^{2}}} y,\qquad L \rightarrow \sqrt{\frac{\alpha}{3\gamma}} L^2,\label{transfor}
\end{eqnarray}
we observe that the line element (\ref{ansatz}) remains invariant, with the metric function $f(r)$ now adopting the following form:
\begin{eqnarray}
&&f(r)= 1-\left(\frac{r}{r_{h}}\right)^{4}.\,\label{L16}
\end{eqnarray} 
Here, we can see that the metric function only has one integration constant, without additional charges. Nevertheless, it is possible to perform a geometry-independent treatment via a probe approximation. In this case, the equations of motion for the electromagnetic field can be solved independently, allowing us to explore a finite charge and density in five dimensions. For higher-dimensional scenarios, see Ref. \cite{Ghotbabadi:2021mus}. From eqs. (\ref{1}) and (\ref{eq:Maxwell})-(\ref{eq:L2FF}), we consider that $V(M)$ reads
\begin{equation}
V(M)=(^{*}M_{\mu \nu} M^{\mu \nu})^2=[^{\ast} (M \wedge M)]^2,\label{L4.1}\\ \end{equation}
where $(^{*})$ is the Hodge star operator. The field equations in the probe approximation, that is $e^{2}\to +\infty$ and $\lambda\to 0$, are given by
\begin{eqnarray}
\nabla^{\mu} \left(F_{\mu \nu}+\frac{\lambda^2}{4}\,M_{\mu\nu} \right) &=& 0,\label{eom1}\\
\nabla^{\tau} (dM)_{\tau \mu \nu} - m^2 M_{\mu \nu} - J(^{\ast}M_{\tau \sigma}M^{\tau \sigma})(^{\ast}M_{\mu \nu}) - F_{\mu \nu} &=&0\,. \label{eom2}
\end{eqnarray}
As we are focusing on the probe limit approximation, we are going to disregard any backreaction coming from the two-form field $M_{\mu \nu}$. To analyze the holographic transport and magnetizing plasma, using the fluid/gravity duality, we consider the gauge fields $M_{\mu \nu}$ and $A_{\mu}$ in the following form \cite{Cai:2015bsa,Cai:2014oca}:
\begin{eqnarray} 
M_{\mu \nu} &=& -p(r)\,dt\wedge dr + \rho(r)\,dx\wedge dy,\label{Mansatz}  \\
A_{\mu} &=& A_{t}(r)\,dt + B x\,dy, \quad F_{\mu \nu} = \partial_{\mu} A_{\nu}-\partial_{\nu} A_{\mu}.\label{Fansatz}
\end{eqnarray}
Here, $B$ is a constant that represents the external magnetic field. With all this information, via eqs. \eqref{ansatz}, \eqref{Mansatz}-\eqref{Fansatz} in the background (\ref{L16}), the field equations \eqref{eom1} and \eqref{eom2} are obtained explicitly, and for the sake of completeness are present in the Appendix \ref{AB}. Given that we are working on probe approximation, we can disregard the back reaction. As the system exhibits asymptotic AdS$_{5}$ behavior, in approaching to the boundary (this is $r \to 0$), we can solve the field equations (\ref{L18})-(\ref{L20}). The solutions in this asymptotic regime are outlined below:
\begin{eqnarray}
&&A_{t}(r)\sim\mu-\sigma r,\\
&&p(r)\sim \frac{4\sigma}{\lambda^{2}}(1+r),\\
&&\rho(r)\sim\rho_{+}r^{\Delta_{+}}+\rho_{-}r^{\Delta_{-}}+\frac{B}{m^2},\label{eq:rho}\\
&&\Delta_{\pm}=\pm\,2mL.\label{eq:m}
\end{eqnarray}
{Here,} $\rho_{+}$ and $\rho_{-}$ are integration constants representing the source and the vacuum expectation value of the dual operator in the boundary field theory (up to a normalization factor) respectively, where to obtain condensation spontaneously, one should take $\rho_{+}=0$, \cite{Cai:2015bsa}. To simplify our calculations, from eq. (\ref{eq:rho}) the integration constants can be defined as $\rho_{+}:= r^{-\Delta_{+}}_{h},\, \rho_{-}:= r^{-\Delta_{-}}_{h}$, and $\rho(r)$ acquires the structure:
\begin{eqnarray}
\rho(r)\sim \left(\frac{r}{r_{h}}\right)^{\Delta_{+}}+\left(\frac{r}{r_{h}}\right)^{\Delta_{-}}-\frac{B}{m^2}.
\end{eqnarray}
Beyond these conditions to the bulk side, we apply Neumann boundary conditions NBC to extract the ratio $\rho/B$. For this, we assume that $Q$ is parameterized through the equation $y=y_{Q}(r)$, analyzing the influence of the Horndeski Lagrangian (\ref{eq:Lhorn}). Together with the above, the induced metric on this surface reads
\begin{eqnarray}
ds^{2}_{\rm ind}=\frac{L^{2}}{r^{2}}\left(-f(r)dt^{2}+dx^{2}+dw^{2}+\frac{g^{2}(r)dr^{2}}{f(r)}\right),\label{Q1} 
\end{eqnarray}
where $g^{2}(r)=1+{y'}^{2}(r)f(r)$ and, as before, $({'})$ denotes the derivative with respect to $r$. The normal vectors on $Q$ are 
\begin{eqnarray}
n^{\mu}=\frac{r}{Lg(r)}\, \Big(0,0,0,\, 1, \, -{f(r)y{'}(r)}\Big),\label{Q2}
\end{eqnarray}
and through the field equation ${\cal F}_{\phi}=0$ (\ref{L12}), one can solve the eq. \eqref{L7} (with ${\cal S}^{Q}_{\alpha\beta}=0$), yielding
\begin{eqnarray}
y{'}(r)&=&\frac{(\Sigma L)}{\sqrt{4-\dfrac{\xi L^{2}}{2r^{2}f(r)}-(\Sigma L)^{2}f(r)}}.\label{prof}
\end{eqnarray}
Here, $\xi$ was given previously in (\ref{eq:xi}), $f(r)$ is the metric function (\ref{L16}), and $\Sigma L=\cos (\theta')$, where $\theta'$ represents the angle between the positive direction of the $y$ axis and $Q$.

\begin{figure}[h!]
\begin{center}
\includegraphics[scale=0.8]{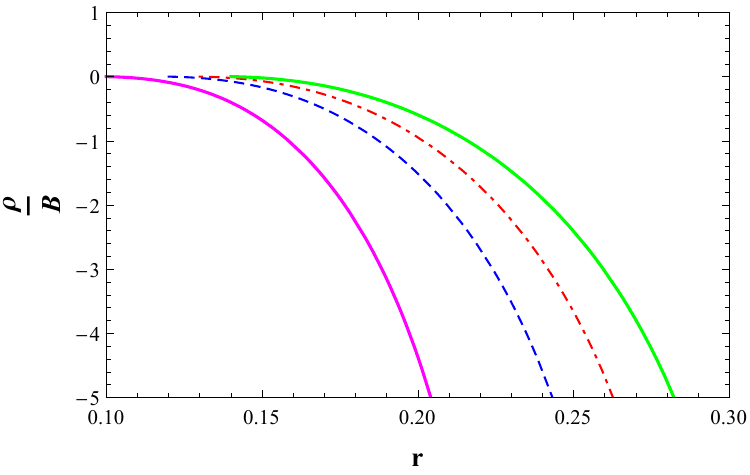}
\caption{Graphic of the $\rho/B$ ratio versus $r$ for different values of the Horndeski parameter $\gamma$. For our analysis, we consider $r_{h}=0.1$, $L=1$, $\theta{'}=2\pi/3$, $\Lambda=-1$, $\alpha=0.5$, $m=1$, and $\gamma=0$ ({\sl pink curve}), $\gamma=0.1$ ({\sl blue dashed curve }), $\gamma=0.2$ ({\sl red dot dashed} curve), and $\gamma=0.3$ ({\sl green thick curve}).}
\label{p0}
\label{ylinhaz}
\end{center}
\end{figure}

Through the steps from Refs. \cite{Melnikov:2012tb,dosSantos:2022scy,Santos:2023flb}, the NBC on the gauge field is $n^{\mu}{F}_{\mu\nu}|_{Q}=0$, while that $B=\sigma$. As the four-dimensional situation, the holographic model (AdS$_{5}$/BCFT$_{4}$) predicts that a constant boundary current in the bulk induces a constant current on the boundary $Q$ \cite{Santos:2023flb}. Furthermore, $n^{\mu}M_{\mu\nu}|_{Q}=0$ provide  
\begin{eqnarray}\label{eq:rho/B}
\frac{\rho(r)}{B}=\dfrac{f(r)y'(r)}{m^{2}},
\end{eqnarray}
where the density $\rho$ and the magnetic field $B$ are dependent on the values of the Horndeski parameters and the polarization tensor. It is interesting to note that the $\rho/B$ ratio is the Hall conductivity, which resembles the quantum Hall effect (QHE). The ratio $\rho/B$ presented in eq. (\ref{eq:rho/B}) is shown in Fig. \ref{p0}, where on the boundary $Q$ the curves of solutions in the plane ($\rho,B$) correspond to a localized condensate \cite{Hartnoll:2008kx,Hartnoll:2009sz}. In addition to the above numerical solution, we can analyze some particular cases regarding the study of the ultraviolet (UV) regimes. Thus, {for the first case,} performing an expansion at $r\to 0$ with, as before, $\Sigma L=\cos(\theta{'})$, the equation (\ref{prof}) becomes 
{\begin{eqnarray}
y_{_{UV}}(r)=y_{0}+ \sqrt{\frac{2}{-\xi L^{2}}}\,{r\cos(\theta{'})},
\end{eqnarray}
where $y_0$ is an integration constant.} In the above equation, considering $\xi\to-\infty$, we have that $y_{_{UV}}(r)=y_{0}={\rm constant}$, which is equivalent to keeping $\xi$ finite together with a zero tension limit $\Sigma\to 0$, {considering the cases $\theta'=\pi/2$ and $\theta'=3\pi/2$. For this regime, we have that the $\rho/B$ ratio takes the form
\begin{eqnarray}
\frac{\rho}{B}=\sqrt{\frac{2}{-\xi L^{2}}}\dfrac{\cos(\theta{'})}{m^{2}}.\label{ratio}
\end{eqnarray}
We can note that although the above result is for five dimensions, is a consistent generalization of a known AdS$_{4}$/CFT$_{3}$ solution, given by the four-dimensional AdS BH with plane symmetry, where it allows only stress-free RS branes in the construction \cite{Melnikov:2012tb,Santos:2023flb}.} Furthermore, we have that uniform static charge density must be supported by a magnetic field. The ratio (\ref{ratio}) illustrates how UV physics is captured; we clarify that this is not an attempt to capture asymptotic freedom with this boundary condition for a UV correspondence to QCD \cite{Gubser:2008yx}. Our prescription shows that asymptotic freedom is replaced by conformal invariance, which occurs in the high-temperature regime and is computed through holographic renormalization.

\section{Free energy and thermodynamic quantities}\label{sec:HR}
To describe both the thermodynamic and hydrodynamic coefficients of a conformal fluid in the presence of a magnetic field, in this section we will calculate the Euclidean on-shell action, which is related to the free energy of the corresponding thermodynamic system.  Let us start with the Euclidean action given by $I_{E}=I_{bulk}+2I_{bdry}$, where
\begin{eqnarray}
I_{bulk}&=&-\frac{1}{16\pi G_N}\int_{\mathcal{N}}d^{5}x \sqrt{g}\Big(R-2\Lambda+\frac{\gamma}{2} G_{\mu \nu} \nabla^{\mu} \phi \nabla^{\nu} \phi\Big) \label{eq:Ibulk}\\
&-&\frac{1}{8\pi G_N}\int_{\mathcal{M}}d^{4}x\sqrt{\bar{\gamma}} \Big(K^{({\bar{\gamma}})}-\Sigma^{(\bar{\gamma})}-\frac{\gamma}{4}(\nabla_{\mu}\phi\nabla_{\nu}\phi n^{\mu}n^{\nu}-(\nabla\phi)^{2})K^{(\bar{\gamma})}-\frac{\gamma}{4}\nabla^{\mu}\phi\nabla^{\nu}\phi K^{(\bar{\gamma})}_{\mu\nu}\Big). \nonumber
\end{eqnarray}
In eq. (\ref{eq:Ibulk}),  $g$ is the determinant of the metric $g_{\mu\nu}$ on the bulk $\mathcal{N}$, $\bar{\gamma}$ is the induced metric, the surface tension (resp. extrinsic curvature) on $\mathcal{M}$ is represented by $\Sigma^{(\bar{\gamma})}$ (resp. $K^{({\bar{\gamma}})}$). The boundary side is represented by $I_{bdry}$, which reads
\begin{eqnarray}
I_{bdry}&=&-\frac{1}{16\pi G_{N}}\int_{\mathcal{N}}d^{5}x \sqrt{g} \left(R-2\Lambda+\frac{\gamma}{2} G_{\mu \nu} \nabla^{\mu} \phi \nabla^{\nu} \phi\right)\label{eq:Lbndy}\\
&-&\frac{1}{8\pi G_N}\int_{Q}d^{4}x\sqrt{h} \Big(
(K-\Sigma)-\frac{\gamma}{4}(\nabla_{\mu}\phi\nabla_{\nu}\phi n^{\mu}n^{\nu}-(\nabla\phi)^{2})K-\frac{\gamma}{4}\nabla^{\mu}\phi\nabla^{\nu}\phi K_{\mu\nu}\Big).\nonumber
\end{eqnarray}
To construct the explicit expression for the bulk action $I_{bulk}$ (\ref{eq:Ibulk}), we need to consider the induced metric on the bulk, which is obtained from the metric ansatz (\ref{ansatz}) after the transformation $\tau=i t$, given by
\begin{eqnarray}
ds^{2}_{ind}=\bar{\gamma}_{\mu\nu}dx^{\mu}dx^{\nu}=\frac{L^{2}}{r^{2}}\left(f(r)d\tau^{2}+dx^{2}+dy^{2}+dw^{2}+\frac{dr^{2}}{f(r)}\right),\label{mett1}
\end{eqnarray}
where  $0 \leq \tau \leq \beta$ with
\begin{eqnarray}\label{eq:Th}
\beta=\frac{1}{T}=\left(\frac{|f'(r)|}{4 \pi}\Big{|}_{r=r_h}\right)^{-1}=\pi\,r_{h}.
\end{eqnarray}
Here, $T$ is the Hawking Temperature, obtained from eqs. (\ref{ansatz}) and (\ref{L16}) respectively. Now, using these elements, we can construct the bulk action $I_{bulk}$. For this part of the process of holographic renormalization, to remove IR diverges in the bulk side, we introduce a cutoff $\epsilon$, which is
\begin{eqnarray}
I_{bulk}&=&\frac{1}{16\pi G_{N}}\int{d^{3}x}\int^{\beta}_{0}{d\tau}\int^{r_{h}}_{\epsilon}{dr\sqrt{g}\left(R-2\Lambda+\frac{\gamma}{2}G^{rr}\psi^{2}(r)\right)}+\nonumber\\
&+&\frac{1}{16\pi G_{N}}\int{d^{3}x}\int^{\beta}_{0}{d\tau}{\frac{L^{2}\sqrt{f(\epsilon)}}{\epsilon^{4}}}=-\frac{L^{2}V}{8r^{3}_{h}G}\left(1-\frac{\xi}{4}\right),\label{eq:bulk}
\end{eqnarray}
with $\xi$ given previously in (\ref{eq:xi}) and, in our notations, $V=\int{d^{3}x}=\Delta{x}\Delta{y}\Delta{w}=(x_2-x_1)(y_2-y_1)(w_2-w_1)$.
Now, computing the $I_{bdry}$, we introduce a cutoff $\epsilon$ to remove the UV divergence on the boundary side, which reads
\begin{eqnarray}
I_{bdry}=\frac{r_{h}L^{2}{\Delta y_{Q}}}{2G_{N}}\left(1-\frac{\xi}{4}\right)\int^{r_{h}}_{\epsilon}{\frac{\Delta y_{Q}(r)}{r^{5}}dr}-\frac{r_{h}L^{2}\sec(\theta{'}){\Delta y_{Q}}}{2G_{N}}\int^{r_{h}}_{\epsilon}{\frac{\Delta y_{Q}(r)}{r^{4}}dr}.\label{Idry}
\end{eqnarray}
In the above equation, $\Delta y_{Q}$ is a constant, and $\Delta y_{Q}(r):=y_{Q}(r)-y_0$ is obtained from the eq. (\ref{prof}) via an approximate analytical solution for
$y(r)$, obtained by performing an expansion for $\xi$ very small and after integration with respect to $r$ (see Refs. \cite{Santos:2023flb,Santos:2021orr} for more details). From the overview point of AdS/CFT correspondence, IR divergences in AdS correspond to UV divergences in CFT where this relation is known as the IR-UV connection. Thus, based on this duality, we can reduce eq. (\ref{Idry}) to the following form:
\begin{eqnarray}
I_{bdry}&=&-\frac{L^{2}\Delta\,y_{Q}}{2 G_{N}}\left(1-\frac{\xi}{4}\right)\left(\frac{\xi\,L^{2}b(\theta{'})}{5r^{3}_{h}}-\frac{q(\theta^{'})}{3r_{h}}\right)\nonumber\\
&+&\frac{L^{2}\sec(\theta{'})\Delta\,y_{Q}}{2G_{N}}\left(\frac{\xi\,L^{2}b(\theta{'})}{5r^{2}_{h}}-\frac{q(\theta^{'})}{2}\right),\label{eq:bdry1}
\end{eqnarray}
\noindent where
\begin{eqnarray}
b(\theta{'})=\frac{\cos(\theta{'})}{\sqrt{2}(7-\cos^{2}(\theta{'}))^{3/2}},\qquad  
q(\theta{'})=\frac{\sqrt{2}\cos(\theta{'})}{\sqrt{7-\cos^{2}(\theta{'})}}\,\label{eq:bdry2}. 
\end{eqnarray}
Given the preceding details and through eqs. (\ref{eq:bulk}) and (\ref{eq:bdry1})-(\ref{eq:bdry2}), we can compute $I_{E}=I_{bulk}+2I_{bdry}$ as:
\begin{eqnarray}
&&I_{E}=-\frac{L^{2}V}{8r^{3}_{h}G_N}\left(1-\frac{\xi}{4}\right)-\frac{L^{2}\Delta\,y_{Q}}{G_{N}}\left(1-\frac{\xi}{4}\right)\left(\frac{\xi\,L^{2}b(\theta{'})}{5r^{3}_{h}}-\frac{q(\theta^{'})}{3r_{h}}\right)\nonumber\\
&&+\frac{L^{2}\sec(\theta{'})\Delta\,y_{Q}}{G_{N}}\left(\frac{\xi\,L^{2}b(\theta{'})}{5r^{2}_{h}}-\frac{q(\theta^{'})}{2}\right)\label{freeEBH}.
\end{eqnarray}
In this context, $I_{E}$ represents the approximate analytical expression for the Euclidean action. This equation plays a crucial role in formulating the free energy $\Omega$ and extracting all the thermodynamic quantities of interest of our model, which is defined as
\begin{equation}\label{FE}
 \Omega=\epsilon-TS=T I_E \,,  
\end{equation}
where the entropy and the energy density, denoted as $S$ and $\epsilon$ respectively, read
\begin{eqnarray}
S=-\frac{\partial\,\Omega}{\partial T}, \qquad \epsilon=\Omega-T \left(\frac{\partial\,\Omega}{\partial T}\right),  \label{BT7}
\end{eqnarray}
and $T$ is, as before, the Hawking temperature. By plugging the Euclidean \textit{on-shell action} $I_E$ from eq. \eqref{freeEBH}, and after expressing the location of the event horizon $r_h$ in function of $T$ via eq. (\ref{eq:Th}), we have that entropy $S$ can be split into two parts:
\begin{eqnarray}\label{eq:ent-total}
S&=&S_{\rm bulk}+S_{\rm bdry},
\end{eqnarray}
where
\begin{eqnarray}
S_{\rm bulk}&=&\frac{L^{2}V}{4r^{3}_{h}G_{N}}\left(1-\frac{\xi}{4}\right),\label{eq:Entbulk}\\
S_{\rm bdry}&=&\frac{L^{2}\Delta\,y_{Q}}{G_{N}}\left(1-\frac{\xi}{4}\right)\left(\frac{\xi\,L^{2}b(\theta{'})}{5r^{3}_{h}}-\frac{q(\theta^{'})}{3r_{h}}\right)\nonumber\\
&-&\frac{L^{2}\sec(\theta{'})\Delta\,y_{Q}}{G_{N}}\left(\frac{\xi\,L^{2}b(\theta{'})}{5r^{2}_{h}}-\frac{q(\theta^{'})}{2}\right).\label{BT8}
\end{eqnarray}
The meaning behind this overall entropy (\ref{BT8}) aligns with the Bekenstein-Hawking expression 
\begin{eqnarray}
S_{BH}=\frac{A}{4G_{N}}\label{BT9}\,,
\end{eqnarray}
where, in this case, the total area $A$ reads
\begin{eqnarray}
A&=&\frac{L^{2}V}{2r^{3}_{h}}\left(1-\frac{\xi}{4}\right)+4L^{2}\Delta\,y_{Q}\left(1-\frac{\xi}{4}\right)\left(\frac{\xi\,L^{2}b(\theta{'})}{5r^{3}_{h}}-\frac{q(\theta^{'})}{3r_{h}}\right)\nonumber\\
&-&4L^{2}\sec(\theta{'})\Delta\,y_{Q}\left(\frac{\xi\,L^{2}b(\theta{'})}{5r^{2}_{h}}-\frac{q(\theta^{'})}{2}\right),\,\label{BT10}
\end{eqnarray}
allowing us to obtain new contribution terms. For the sake of completeness, for the entropy $S$, we have that the information storage (delimited via the area $A$) increases as the magnitude of $\xi$ increases, as long as $\xi<0$. Additionally, from eq. (\ref{BT10}) for low temperatures ($T\to 0$), we have that 
\begin{eqnarray}
S^{res}_{bdry}=\frac{L^{2}\sec(\theta{'})\Delta\,y_{Q}q(\theta{'})}{2G_{N}}.\label{resid}
\end{eqnarray}
As we will see in the following section, this residual information, present in the entropy, for this five-dimensional\footnote{For more discussions in three and four dimensions, see Refs.\cite{Santos:2021orr,Santos:2023flb}.} scenario will be responsible for a significant increase in both $\zeta/S$ and $\eta/S$, and also lead to deviations in the value of $c^{2}_{s}$. Moreover, in conjunction with the Horndeski gravity parameters, this residual information will capture the impact of the magnetic field on the conformal plasma through the boundary conditions that resulted in the ratio $\rho/B$, which was defined in eq. (\ref{ratio}). Thus, it is convenient to write $S^{res}_{bdry}$ in terms of the magnetic field, which leads us to $S^{magnetic}_{bdry}$, which is expressed as follows:
\begin{eqnarray}
S^{magnetic}_{bdry}&=&\frac{L^{2}\Delta\,y_{Q}}{G_{N}}\left(1-\frac{\xi}{4}\right)\left(-\frac{2B^{2}\cos^{2}(\theta^{'})}{m^{2}\rho^{2}}\frac{b(\theta{'})}{5r^{3}_{h}}+\frac{q(\theta^{'})}{3r_{h}}\right)\nonumber\\
&-&\frac{L^{2}\sec(\theta{'})\Delta\,y_{Q}}{G_{N}}\left(-\frac{2B^{2}\cos^{2}(\theta^{'})}{m^{2}\rho^{2}}\frac{b(\theta{'})}{5r^{2}_{h}}+\frac{q(\theta^{'})}{2}\right),\label{BT11ext}
\end{eqnarray}
Here, $S^{magnetic}_{bdry}$ is the entropy bound. The charge density contributed to the magnetized plasma must be finite in the probe limit.

\section{Bulk viscosity, shear viscosity and speed of sound }\label{sec:hol}

In this section, we delve into the study of crucial quantities within the context of the QGP: $\zeta/S$, $\eta/S$, and $c_s^2 = S/C_V$ (see for example Refs. \cite{Buchel:2007mf,Kharzeev:2007wb,Karsch:2007jc,Gubser:2008yx}). Here, $C_V$ represents the heat capacity, and $S$ denotes the entropy density.

The QGP, a state of matter prevalent in the early universe, is recreated in heavy ion collision experiments \cite{Kharzeev:2007jp,Fukushima:2008xe}. The  $\zeta/S$ ratio plays a pivotal role in maintaining the equilibrium of a fluid experiencing slight expansion or compression. On the other hand, the $\eta/S$ ratio quantifies the energy dissipation resulting from relative movements among fluid layers \cite{Torres-Rincon:2012sda,Mishra:2020onx}. Furthermore, the square of the speed of sound $c_s^2$ characterizes the velocity at which disturbances propagate through the medium \cite{Bluhm:2007nu}. Calculating the bulk viscosity of strongly coupled thermal plasmas, which is dual to supergravity backgrounds supported by a scalar field, necessitates determining the primary dissipative term in the dispersion relation for sound waves \cite{Buchel:2007mf,Gubser:2008yx,Kharzeev:2007wb,Karsch:2007jc,Kharzeev:2007jp,Fukushima:2008xe,Endrodi:2018ikq}.

\begin{figure}[h!]
\begin{center}
\includegraphics[scale=0.65]{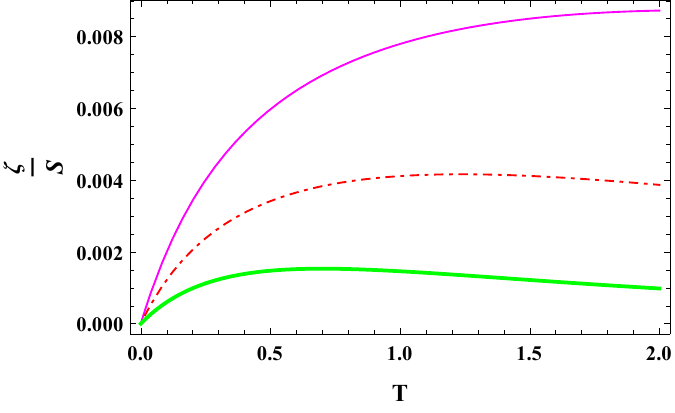} \, \includegraphics[scale=0.65]{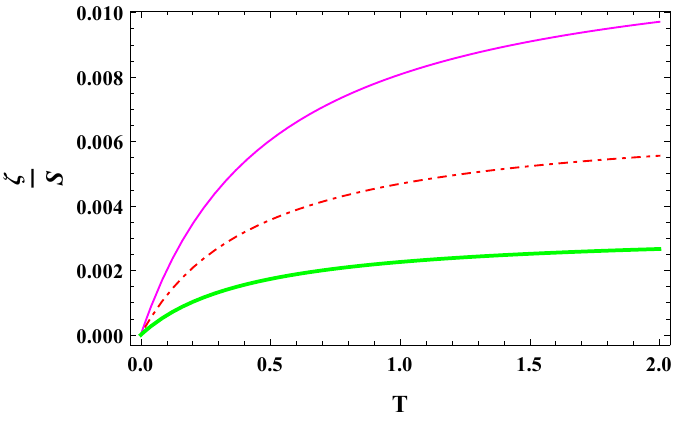}
\caption{{\bf Left panel:} The behavior of the $\zeta/S$ ratio as a function of the temperature $T$ (where for our case we consider $B=(4/5)T$) for different values for $\alpha=8/3$, $\rho=1/4$, $\Lambda=-1$, $\gamma=1$ ({\sl pink curve}), $\gamma=2$ ({\sl red dot dashed curve}), $\gamma=2.5$ ({\sl green thick curve}). {\bf Right panel:} The behavior of $\zeta/s$ for $B=0$ considering the same values showed previously.}\label{zeta1}
\end{center}
\end{figure}

\begin{figure}[h!]
\begin{center}
\includegraphics[scale=0.65]{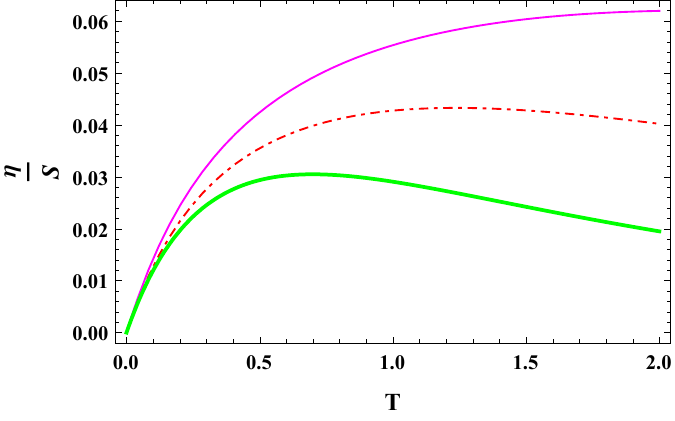} \, \includegraphics[scale=0.65]{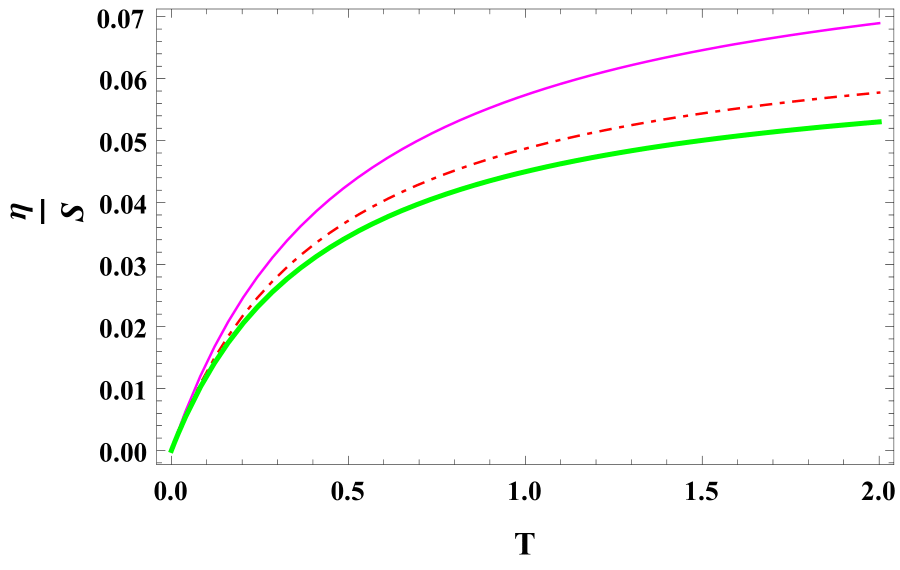}
\caption{{\bf Left panel:} The behavior of the $\eta/S$ ratio as a function of the temperature $T$ for different values for $\alpha=8/3$, $B=(4/5)T$, $\rho=1/4$, $\Lambda=-1$, $\gamma=1$ ({\sl pink curve}), $\gamma=2$ ({\sl red dot dashed curve}), $\gamma=2.5$ ({\sl green thick curve}). {\bf Right panel:} The behavior of $\eta/s$ for $B=0$ considering the same values showed previously.}\label{visco1}
\end{center}
\end{figure}

\begin{figure}[h!]
\begin{center}
\includegraphics[scale=0.7]{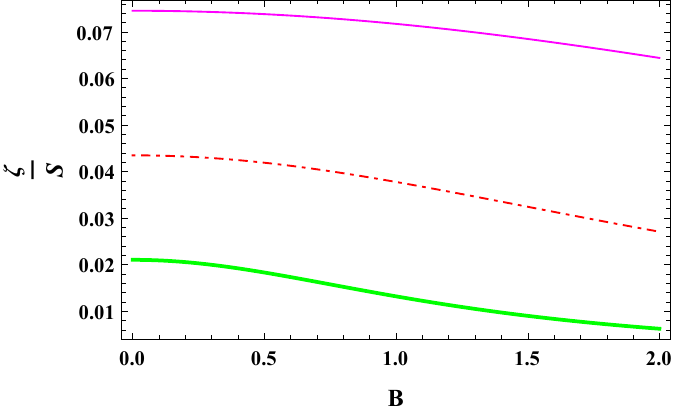} \,\includegraphics[scale=0.7]{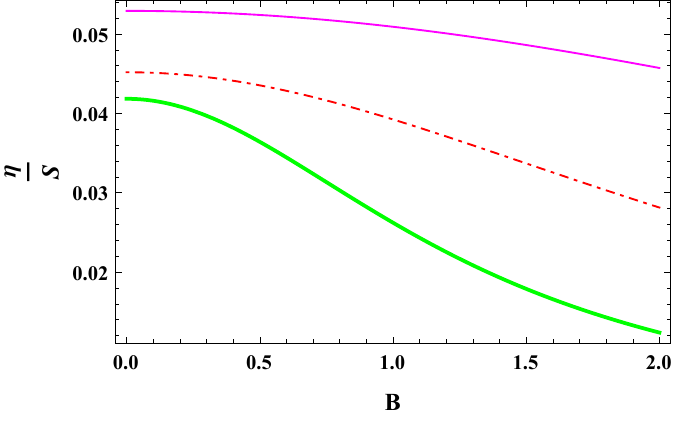}
\caption{{\bf Left panel:} The behavior of $\zeta/S$ with respect to the magnetic field $B$. {\bf Right panel:} The behavior of $\eta/S$ with respect to the magnetic field $B$. For both situations, we consider $\alpha=8/3$, $T=4/5$, $\rho=1/4$, $\Lambda=-1$, $\gamma=1$ ({\sl pink curve}), $\gamma=2$ ({\sl red dot dashed curve}) and $\gamma=2.5$ ({\sl green thick curve}) respectively.}\label{zetavisco2}
\end{center}
\end{figure}

\begin{figure}[h!]
\begin{center}
\includegraphics[scale=0.7]{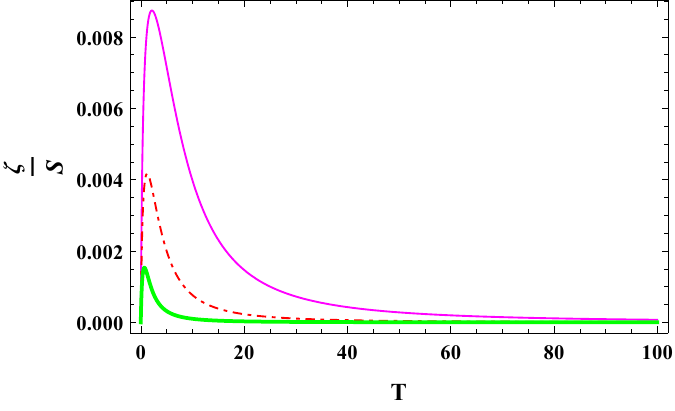} \includegraphics[scale=0.7]{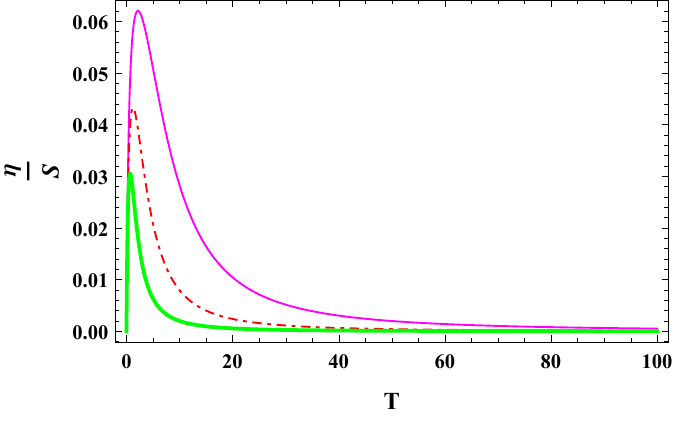}
\caption{{\bf Left panel:} the behavior of $\zeta/S$ with respect to $T$. {\bf Right panel:} The behavior of $\eta/S$ v/s $T$. For both situations, we consider $\alpha=8/3$, $B=4/5$, $\rho=1/4$, $\Lambda=-1$, $\gamma=1$ ({\sl pink curve}), $\gamma=2$ ({\sl red dot dashed curve}) and $\gamma=2.5$ ({\sl green thick curve}).}\label{zetaeta}
\end{center}
\end{figure}

In our work, we propose a methodology applicable to the Horndeski scenario, requiring solely the primary frequency dependence of a suitable signal. This approach facilitates the calculation of Green's Function in the low-frequency limit \cite{Santos:2023flb,Feng:2015oea}, extracting $\zeta/S$ and $\eta/S$ ratios,  given by
\begin{eqnarray}
&&\frac{\zeta}{S}=\frac{\sqrt{3}}{24\pi\mathcal{F}}\sqrt{\frac{\alpha+\gamma\Lambda}{\alpha-\gamma\Lambda}}\label{bulk},\\
&&\frac{\eta}{S}=\frac{1}{4\pi\mathcal{F}}\sqrt{\frac{3\alpha+\gamma\Lambda}{\alpha-\gamma\Lambda}},\label{visc}
\end{eqnarray}
where all the details about the computations are presented in Appendix \ref{viscosity}. In the low-temperature stages of the QGP for realistic heavy ion collisions, we note the presence of the magnetic field strength on both the $\zeta/S$ and $\eta/S$ graphs across various temperature ranges (see Figs. \ref{zeta1} and \ref{visco1}, left panel) \cite{ALICE:2019sgg,STAR:2019clv}. Additionally, when the temperature $T$ is held constant, this effect is further elucidated (see Figure \ref{zetavisco2}). However, when the magnetic field is turned off ($B=0$ and Figs. \ref{zeta1}-\ref{visco1}, right panel), the Horndeski parameters $\alpha$ and $\gamma$ produce conformal symmetry breaking at low temperatures, but at high temperatures, the conformal symmetry is recovered as predicted by QCD. The above is shown for a wider range of values for $T$ (see Fig. \ref{zetaeta}).

It is interesting to note that with eqs. (\ref{bulk})-(\ref{visc}) we can consider the $\zeta/\eta$ ratio, which presents a similar expression to those presented in \cite{Gubser:2008sz,Buchel:2007mf}, but now a distinctive feature is that:
\begin{eqnarray}
\frac{\zeta}{\eta}=\frac{1}{6}\sqrt{\frac{3(\alpha+\gamma\Lambda)}{3\alpha+\gamma\Lambda}},\label{buchelb}
\end{eqnarray}
where $\alpha=-\gamma\Lambda$ does not form a condensate, implying $\zeta=0$. This outcome aligns with the prediction in \cite{Gubser:2008sz}, implying no contribution of bulk viscosity ($\zeta_{BCFT}\to0$) on the BCFT side, {this regime indicates that the boundary stress-energy tensors, denoted as $T^{Q}_{\alpha\beta}$, residing in $Q$, must be null trace. However, to check this discussion, we need to compute the trace of $T^{Q}_{\alpha\beta}$; for more details, see the next section}.

The expression provided in eq. (\ref{buchelb}) indicates that during the plasma phase of the fluid, the scalar field vanishes, leading to the absence of bulk viscosity in the fluid. Consequently, it is intriguing to observe the transition from a fluid characterized by low shear viscosity to a plasma state exhibiting low bulk viscosity, regardless of the presence of an external magnetic field. Furthermore, we note that the above represents a plasma state composed of quarks and gluons, as illustrated in \cite{Ballon-Bayona:2022uyy}, wherein only vector bosons are present and scalars are absent.

As depicted in Fig. \ref{zetaeta}, with increasing temperature, both $\eta/S$ and $\zeta/S$ tend towards zero. Furthermore, in the limit as $\gamma$ approaches zero, we recover the result of the Chamblin-Reall background in five dimensions 
$$\frac{\zeta}{\eta}=\frac{1}{6},$$
as discussed in \cite{Gubser:2008sz}. Hence, our results exhibit remarkable consistency when we exclude the Horndeski contributions, controlled by the $\gamma$ parameter. 

Additionally, in Fig. \ref{zetaeta} the curves represent the boundary contributions stemming from the residual boundary entropy, akin to the Buchel bound. In the theory we are working with, this bound is described by eq. (\ref{buchelb}). 

Finally, we present the square of the speed of sound $c^{2}_{s}$, defined as:
\begin{eqnarray}
    c_s^2 := \frac{\partial p}{\partial \epsilon}
    =  \left(\frac{\partial p}{\partial T}\right) \left( \frac{\partial T}{\partial \epsilon}\right),
\end{eqnarray}
where $p$ is the pressure of a
system and $\epsilon$ is its energy density. For the case $c^{2}_{s} \approx \frac{1}{3}$, we observe the expected behavior for a non-interacting gas of massless partons \cite{Bluhm:2007nu,Karsch}.  Identifying: 
\begin{eqnarray}
    \frac{\partial T}{\partial \epsilon} = \left(\frac{\partial \epsilon}{\partial T}\right)^{-1} = C_V^{-1} \,,\qquad 
    \frac{\partial p}{\partial T} = S\,, 
\end{eqnarray}
we have\footnote{It is also very common to describe the sound speed as $c^2_s = \frac{\partial(\ln{T})}{\partial(\ln{S})}$, where, as before, $T$ is the Hawking temperature and $S$ is the entropy \cite{Santos:2021orr}.} 
\begin{eqnarray}
    c_s^2 &=& \frac{S}{C_V};\quad\,C_{V}=T\bigg(\frac{\partial S}{\partial T}\bigg)_{V,B}=-T\bigg(\frac{\partial^{2} \Omega}{\partial T^{2}}\bigg)_{V,B}.
\end{eqnarray}

\begin{figure}[h!]
\begin{center}
\includegraphics[scale=0.7]{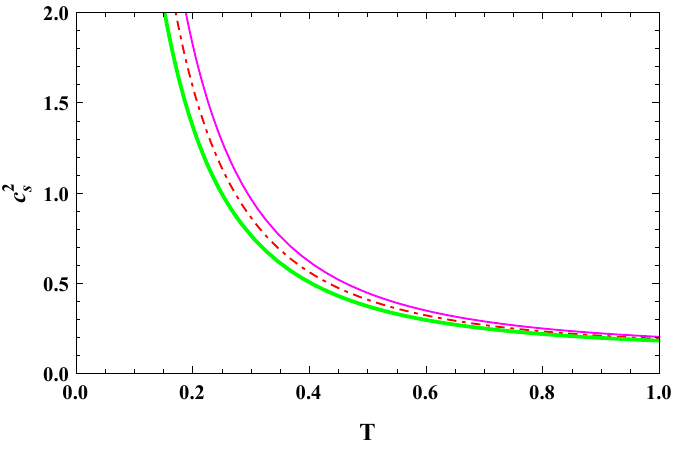}\,\includegraphics[scale=0.7]{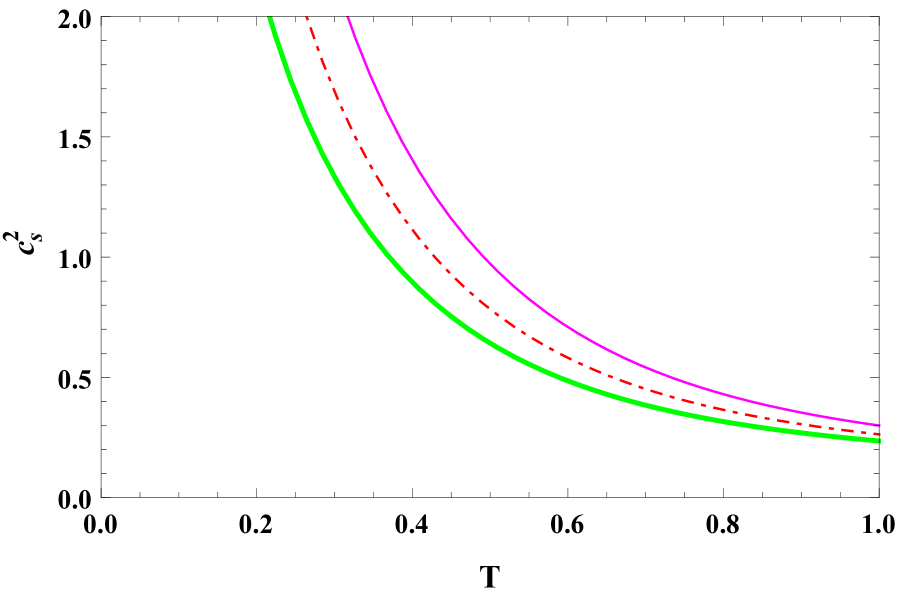}
\caption{{\bf Left panel:} The behavior of  $c^{2}_{s}$ versus the temperature $T$ (considering for this case $B=(4/5)T$). {\bf Right panel:} The behavior of $c^{2}_{s}$ with respect to the temperature $T$ for $B=0$. For both situations, we have considered: $\alpha=8/3$, $m=1/8$, $\rho=1/4$, $\Lambda=-1$, $\theta{'}=2\pi/3$ with $\gamma=1$ ({\sl pink curve}), $\gamma=4$ ({\sl red dot dashed  curve}), $\gamma=8$ ({\sl green thick  curve}).}\label{ssvt}
\end{center}
\end{figure}

In Fig. \ref{ssvt}, we present the behavior of the square of the speed of sound $c^2_s$ versus the temperature $T$. Here it is important to note that $c_s^2$ in the left panel deviates from the value $1/3$ due to the anisotropy acquired from the magnetic field and the Horndeski parameters. The right panel displays $c_s^2$ for zero magnetic fields ($B=0$). 

In the absence of a magnetic field, the system's phase aligns with the Horndeski gravity phase for small values and high temperatures $T$. For instance, $c^{2}_{s}=1/3$ for $\gamma=1$ ({\sl pink curve}). However, for larger values of $\gamma$ and $T$, the residual term of the entropy linked with the boundary $Q$ deviates the value of $c^{2}_{s}=1/3$. Instead, $c^{2}_{s}$ is approximately $0.26$ for $\gamma=4$ ({\sl red dot-dashed curve}) and $0.24$ for $\gamma=8$ ({\sl green thick curve}), respectively.

\section{Fluid/gravity correspondence: the anisotropic pressures}\label{fluid/grav}

As was shown previously, at the moment to calculate the $\zeta/S$ ratio, the fact that $\zeta$ is not zero implies that the stress-energy tensor for the fluid, as described by the BCFT$_{4}$ side, is anisotropic. This anisotropy is produced due to the presence of the boundary in Horndeski gravity and the magnetic field.  This is characterized by the difference between the pressure along the direction parallel to the magnetic field ($p_{rr}$) and the pressure along the direction transversal to the magnetic field ($p_{xx}=p_{yy}=p_{ww}$). 

In the subsequent analysis, we aim to derive expressions for both the longitudinal ($p_{rr}$) and transverse pressures ($p_{xx}$) in terms of the UV coefficients of the metric. To achieve this, we construct the stress tensor, from the limit field theory in Horndeski gravity \cite{DosSantos:2022exb}, and establish a fluid/gravity correspondence with an external magnetic field within the Horndeski gravity framework.  

Through the renormalization procedure, the form of the stress-energy tensor $T_{\alpha\beta}$ can be written as:
\begin{eqnarray}
&&T_{\alpha\beta}=-\frac{L^{2}}{16\pi G_{N}r^{2}}\left[K_{\alpha\beta}-h_{\alpha\beta}(K-\Sigma)+\frac{\gamma}{4}H_{\alpha\beta}-\kappa T^{R}_{\alpha\beta}-\kappa T^{ct}_{\alpha\beta}\right],\label{T6}\\
&&H_{\alpha\beta}=(\nabla_{\sigma}\phi\nabla_{\rho}\,\phi n^{\sigma}n^{\rho}-(\nabla\phi)^{2})(K_{\alpha\beta}-h_{\alpha\beta}K)-(\nabla_{\alpha}\phi\nabla_{\beta}\phi)K.\label{T7}
\end{eqnarray}
Here, $T^{R}_{\alpha\beta}$ and $T^{ct}_{\alpha\beta}$ are the possible contributions of extrinsic curvature and counter term, respectively, which for our case $T^{R}_{\alpha\beta}=T^{ct}_{\alpha\beta}=0$. According to \cite{Endrodi:2018ikq}, we can write the stress-energy tensor in two parts as follows
\begin{eqnarray*}
&&T_{\alpha\beta}=T^{BCFT}_{\alpha\beta}+T^{magnetic},\\
&&T^{BCFT}_{\alpha\beta}=-\frac{L^{2}}{16\pi G_{N}r^{2}}[K_{\alpha\beta}-h_{\alpha\beta}(K-\Sigma)],\\
&&T^{magnetic}=-\frac{L^{2}}{16\pi G_{N}r^{2}}\left(\frac{\gamma}{4}H_{\alpha\beta}\right)
\end{eqnarray*}
where $T^{magnetic}_{\alpha\beta}$ has the contributions of the magnetic field through DBC via equation (\ref{ratio}). This dependence becomes clear when we extract the energy density $\epsilon$ and pressure $p$. In particular, renormalized quantities relates with the magnetic field that appears in $\epsilon$ and $p$ are due $\rho/B$ ratio, where $\xi^{-1}B^2$ bears resemblance to discussions found in \cite{Endrodi:2018ikq}. Consequently, physical observables remain invariant under this point-like renormalization. Using the induced metric:
\begin{eqnarray}
h_{\alpha\beta}=-\frac{L^{2}f}{r^{2}}dt^{2}+\frac{L^{2}}{r^{2}}dx^{2}+\frac{L^{2}}{r^{2}}dw^{2}+\frac{L^{2}g^{2}}{r^{2}f}dr^{2},
\end{eqnarray}
where the energy density $\epsilon$ and pressure $p$ are given by
\begin{eqnarray}\label{eq:epsilon-p}
\epsilon=u^{\mu}u^{\mu}T_{\mu\mu}, \qquad p=\frac{1}{2}\left(\epsilon+T^{\mu}_{\mu}\right),
\end{eqnarray}
with velocity $u^{\mu}=\frac{dx^{\mu}}{d\tau}$ calculated in the moving frame, and the velocity is given by
\begin{eqnarray}
(u^{t},u^{x},u^{w},u^{r})=\left(\frac{r}{L\sqrt{f(r)}},0,0,0\right),
\end{eqnarray}
we can write eq. (\ref{eq:epsilon-p}) in the following form:
\begin{eqnarray*}
\epsilon&=&\frac{L^{2}}{16\pi G_{N}r^{3}}\left(\Sigma L+\frac{ry^{'}f^{'}+2rfy^{''}-2f^{2}y^{'3}-2y^{'}f}{2(1+y^{'2}f)^{3/2}}\right)+\xi\,L^{2}\mathcal{A},\\
p_{xx}&=&p_{yy}=p_{ww}=-\frac{L^{2}}{16\pi G_{N}r^{3}}\left(\Sigma L-\frac{2(2f-rf^{'})y^{'}+f(4f-rf^{'})y^{'3}-2rfy^{''}}{2(1+y^{'2}f)^{3/2}}\right)\nonumber\\
&+&\frac{\xi\,L^{2}\mathcal{A}}{2},\\
p_{rr}&=&-\frac{L^{2}}{16\pi G_{N}r^{3}}\left(\Sigma L-\frac{4y^{'}f-ry^{'}f^{'}}{2(1+y^{'2}f)^{1/2}}\right)-\frac{\xi\,L^{2}}{2}\mathcal{A},
\end{eqnarray*}
where
\begin{eqnarray*}
\mathcal{A}=\frac{r^{3}y{'}f{'}-2f{'}y{'}-10fy{'}}{2r^{2}fgL}+\frac{g}{r^{4}f^{2}}\left(\frac{1}{2}y{'}f{'}+2y{'}\right)+\frac{rf{'}y{'}-fy{'}+rfy{''}}{2fLg^{3}},
\end{eqnarray*}
and through eq. (\ref{ratio}) we have that 
\begin{eqnarray}
\epsilon&=&\frac{L^{2}}{16\pi G_{N}r^{3}}\left(\Sigma L+\frac{ry^{'}f^{'}+2rfy^{''}-2f^{2}y^{'3}-2y^{'}f}{2(1+y^{'2}f)^{3/2}}\right)-\frac{2\mathcal{A}B^{2}\cos^{2}(\theta{'})}{m^{4}\rho^{2}},\label{dens1}\\
p_{xx}&=&p_{yy}=p_{ww}=-\frac{L^{2}}{16\pi G_{N}r^{3}}\left(\Sigma L-\frac{2(2f-rf^{'})y^{'}+f(4f-rf^{'})y^{'3}-2rfy^{''}}{2(1+y^{'2}f)^{3/2}}\right)\nonumber\\
&-&\frac{\mathcal{A}B^{2}\cos^{2}(\theta{'})}{m^{4}\rho^{2}},\label{dens2}\\
p_{rr}&=&-\frac{L^{2}}{16\pi G_{N}r^{3}}\left(\Sigma L-\frac{4y^{'}f-ry^{'}f^{'}}{2(1+y^{'2}f)^{1/2}}\right)+\frac{\mathcal{A}B^{2}\cos^{2}(\theta{'})}{m^{4}\rho^{2}},\label{dens3}
\end{eqnarray}
showing us the presence of symmetry breaking due to $p_{xx,yy,ww}\neq p_{rr}$, and the space is anisotropic. 

It is established that RS brane solutions accommodate external magnetic fields, as detailed in \cite{Fujita:2012fp}, while solutions involving external electric fields are precluded. Thus, through the discussions of \cite{Ahn:2022azl}, we have the electric polarization vector $P^{\mu}$ and the magnetization $M$, associated with the electric field $E^{\mu}$ and magnetic field $B$, through the susceptibilities denoted as $\chi_{EE}$ and $\chi_{BB}$, are given by
\begin{eqnarray}
&&P^{\mu}=\chi_{EE}E^{\mu},\quad\,M=\chi_{BB}B,\\
&&\chi_{EE}=\dfrac{\partial^2 p_{rr}}{\partial E^{2}},\quad\,\chi_{BB}=\dfrac{\partial^2 p_{rr}}{\partial B^{2}},
\end{eqnarray}
where $\chi_{BB}$ takes the form
\begin{eqnarray}
\chi_{BB}=\frac{\mathcal{A}\cos^{2}(\theta{'})}{m^{4}\rho^{2}}.\label{xibb}
\end{eqnarray}
Here we note that when $\chi_{EE}=0$,  the electric polarization vector vanishes ($P^{\mu}=0$). The external magnetic field does not produce polarization of the magnetic moments of the fluid.

\begin{figure}[h!]
\begin{center}
\includegraphics[scale=0.8]{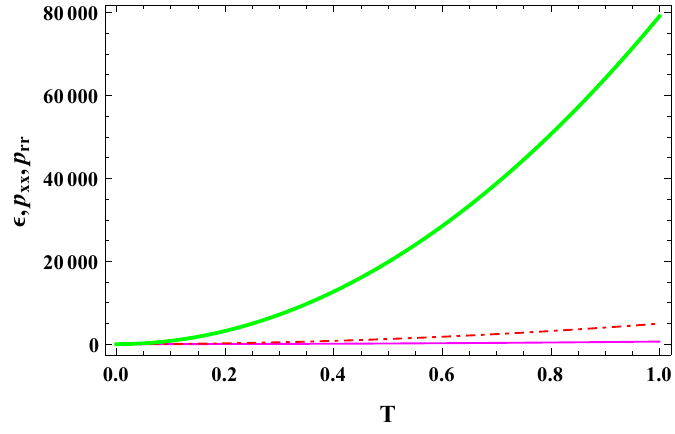}
\caption{The figure presents the behavior of $\epsilon$ (eq. (\ref{rh1})), $p_{xx}$ (eq. (\ref{rh2})), and $p_{rr}$ (eq.(\ref{rh3})), with respect to $T$. Here, we consider  $\alpha=8/3$, $m=1/8$, $B=(4/5)T$, $\rho=1/4$, $\Lambda=-1$ and $\theta{'}=2\pi/3$, where $\epsilon$ is represented via the {\sl pink curve}, $p_{xx}$ through the {\sl red dot dashed curve}, while that $p_{rr}$ is given by the {\sl green thick curve}.}\label{epsilon-p}
\end{center}
\end{figure}

One aspect of the fluid/gravity correspondence is that $T^{Q}_{\alpha\beta}$ offers a natural condition to discriminate between different $Q$ profiles. To extract the fluid profile, we have to impose the condition $p_{xx,yy,ww}=p_{rr}$, producing a magnetized Pascal fluid, which reads
\begin{eqnarray}
\frac{2fy''+f{'}y'}{2r^{3}\sqrt{1+(y')^{2}f}}=0,
\end{eqnarray}
which can be integrated, obtaining $f(y')^{2}=\mbox{constant}$. Thus, the general solution, which yields a fluid-like theory on $Q$, is provided by the profile
\begin{eqnarray}
\Delta y_{Q}(r)=\int^{r}_{0}{\frac{\cot(\theta{'})ds}{\sqrt{f(s)}}}.
\end{eqnarray}
From this way, we have a profile for the tensor $T_{\alpha\beta}$ defined in $Q$, which is consistent with \cite{Magan:2014dwa}. Replacing $f(y')^{2}=\cot^{2}(\theta{'})$ in eqs. (\ref{dens1})-(\ref{dens3}), we have
\begin{eqnarray}
\epsilon&=&\frac{2L^{2}\cos(\theta{'})}{16\pi\,G_{N}r^{3}}(1-\sqrt{f})-2MB,\\
p_{xx}&=&\frac{L^{2}\cos(\theta{'})}{32\pi\,G_{N}r^{3}}\left(\frac{4f-rf{'}-4\sqrt{f}}{\sqrt{f}}\right)-MB,\\
p_{rr}&=&\frac{L^{2}\cos(\theta{'})}{32\pi\,G_{N}r^{3}}\left(\frac{4f-rf{'}-4\sqrt{f}}{\sqrt{f}}\right)+MB,
\end{eqnarray}
where for equation (\ref{xibb}), $\mathcal{A}$ takes the form:
\begin{eqnarray*}
\mathcal{A}&=&\frac{\cot(\theta{'})}{3\sqrt{f}L}\left(\frac{r^{3}f{'}-2f{'}-10f}{r^{2}f\sqrt{1-\cot^{2}(\theta{'})}}+\frac{\sqrt{1-\cot^{2}(\theta{'})}(f{'}+4)}{r^{4}f{'}}+\frac{rf{'}-f
+rff{'}}{(1-\cot^{2}(\theta{'}))^{3/2}}\right),\\
\end{eqnarray*}
where IR divergences in AdS correspond to UV divergences in CFT. Thus, in the UV ($r\to\,0$) regimes, the fluid has a conformal behavior, allowing us to extract the following expressions:
\begin{eqnarray}
&&\epsilon=\frac{2L^{2}\cos(\theta{'})}{16\pi\,G_{N}r^{3}}-\frac{2 \mathcal{A} B^2}{r^{2}},\label{rh1}\\
&&p_{xx}=\frac{L^{2}\cos(\theta{'})}{32\pi\,G_{N}r^{3}}-\frac{\mathcal{A} B^2}{r^{2}},\label{rh2}\\
&&p_{rr}=\frac{L^{2}\cos(\theta{'})}{32\pi\,G_{N}r^{3}}+\frac{\mathcal{A} B^2}{r^{2}},\label{rh3}
\end{eqnarray}
where $\epsilon$, $p_{xx}$ and $p_{rr}$ near the boundary are the expectations value of the $\mathcal{N}=4$ Super Yang-Mills stress-energy tensor (for more discussions see \cite{Endrodi:2018ikq}), while that $\mathcal{A}$ reads
\begin{eqnarray}
\mathcal{A}=-\frac{1}{3(m\,\rho)^{4}}\frac{\cos^{2}(\theta{'})\cot(\theta{'})}{\sqrt{1-\cot^{2}(\theta{'})}}.\label{chim1}
\end{eqnarray}
Now, we can see that the stress tensor trace 
\begin{equation}\label{trace}
\langle T^\alpha_{\ \ \alpha}\rangle=-\epsilon+3p_{xx}+p_{rr},
\end{equation}
disappears, as expected for a conformal fluid, where if $B=0$ we have from eq. (\ref{trace}) that $\epsilon=4p$, which is a conformal behavior \cite{Magan:2014dwa}.  Illustrated in Fig. \ref{epsilon-p}, the discrepancy between longitudinal and transverse pressures in the presence of a magnetic field $B$ is evident. From a phenomenological point of view, these curves for the longitudinal and transverse pressures resemble the results presented in \cite{Bali:2014kia}. Thus, we can conclude that the anisotropy between $p_{rr}$ and $p_{xx}$ increases due to the presence of the magnetic field.

\section{Conclusions and discussions}\label{v4}

In five dimensions, we study fluid thermodynamics employing the holographic framework AdS$_{5}$/BCFT$_{4}$, via a BH in the presence of an external magnetic field, together with a gravity model represented by the Horndeski theory (\ref{eq:Lhorn}). Our study is motivated by the transport coefficients in strongly coupled $\mathcal{N}=4$ Super Yang-Mills plasma, specifically in the presence of a magnetic field. 

A pivotal component of our analysis involves the presentation of the five-dimensional boundary $Q$ profile, where the parameter $\gamma$ from the model (\ref{eq:Lhorn}) takes a providential role. The above allows us to obtain the $\rho/B$ ratio, where the density $\rho$ and the magnetic field $B$ are dependent on the values of the Horndeski parameters and the polarization tensor.

For our analysis, we compute the free energy $\Omega$, allowing us to obtain several thermodynamic quantities, such as the entropy. Here, the introduction of an external magnetic field $B$ aligns with findings presented in \cite{Ahn:2022azl,Ballon-Bayona:2022uyy}. At low temperatures, a residual expression emerges, as described by eq. (\ref{resid}), providing an avenue to explore a significant increase in both $\zeta/S$ and $\eta/S$, and also leading to deviations in the value of $c^{2}_{s}$, placing a special emphasis on the anisotropic effects induced by the magnetic field $B$ as well as by the parameters of the model (\ref{eq:Lhorn}). Our study delves into the thermodynamic properties of the fluid, particularly focusing on its behavior in higher temperatures, resembling a robust plasma within the influence of a magnetic field.

Additionally, the fluid/gravity correspondence with an external magnetic field $B$ is presented. Through the renormalization procedure, we derive the energy density $\epsilon$ and pressure $p$. Here, the presence of an external magnetic field, along with the Horndeski model, induces modifications to the energy density and equation of state, resulting in anisotropy. The above is characterized by the difference between the pressures $p_{rr}$ and $p_{xx}=p_{yy}=p_{ww}$.

The gravity model presented here is based on a specific truncation of the Horndeski theory \cite{Rinaldi:2012vy}. To further explore this model, it would be interesting to extend the thermodynamics and hydrodynamics study to include BHs in arbitrary dimensions, where now the action reads
\begin{eqnarray*}\label{action}
S=\int d^{D}x \sqrt{-g} \left[G_2 +G_4 R+G_{4X}\left(\left(\Box\phi\right)^2
-\left(\nabla_\mu\nabla_\nu\phi\right)^2\right) \right],
\end{eqnarray*}
where $G_2$ and $G_4$ are arbitrary functions of the kinetic term 
$$X=-\frac{1}{2}\nabla_{\mu}\phi \nabla^{\mu} \phi,$$ 
and $G_{4X}=\partial G_4/\partial X$.

Furthermore, we can also consider other models of modified gravity theories that include higher-order Galileon terms (see, for example, Ref. \cite{Nicolis:2008in}). These higher-order terms play a key role in the Vainshtein effect, which ensures that these modified theories remain consistent with observational tests by decoupling the scalar field from matter in gravitationally bound systems. Another essential implication of considering higher-order Galileon terms is the theoretical-phenomenological connection. For instance, in the context of strong interactions in this regime, these modifications in gravity are promising for future work exploring holographic transport with higher-order Galileon terms. Specifically, we will examine the square of the speed of sound, bulk and shear viscosity, and the anisotropic properties induced by the magnetic field in the magnetized conformal plasma.

\begin{acknowledgements}
The authors thank Matteo Baggioli for the fruitful discussions. M.B. is supported by PROYECTO INTERNO UCM-IN-22204, L\'INEA REGULAR. The authors thank the Referee for the comments and suggestions to improve the paper.
\end{acknowledgements}






\appendix

\section{Equations of motion}\label{AA}

In this section, we present the explicit expression for the equations of motion of ${\cal E}_{\mu\nu}$, ${\cal E}_{\phi}$ and ${\cal F}_{\phi}$:

\begin{eqnarray}
{\cal E}_{\mu\nu}&=& G_{\mu\nu}+\Lambda g_{\mu\nu}-\frac{\alpha}{2}\left(\nabla_{\mu}\phi\nabla_{\nu}\phi-\frac{1}{2}g_{\mu\nu}\nabla_{\lambda}\phi\nabla^{\lambda}\phi\right) \label{LEH}\\
                  &+&\frac{\gamma}{2}\left(\frac{1}{2}\nabla_{\mu}\phi\nabla_{\nu}\phi R-2\nabla_{\lambda}\phi\nabla_{(\mu}\phi R^{\lambda}_{\nu)}-\nabla^{\lambda}\phi\nabla^{\rho}\phi R_{\mu\lambda\nu\rho}\right)\nonumber\\
									&+&\frac{\gamma}{2}\left(-(\nabla_{\mu}\nabla^{\lambda}\phi)(\nabla_{\nu}\nabla_{\lambda}\phi)+(\nabla_{\mu}\nabla_{\nu}\phi)\Box\phi+\frac{1}{2}G_{\mu\nu}(\nabla\phi)^{2}\right)\nonumber\\
									&-&\frac{\gamma g_{\mu\nu}}{2}\left(-\frac{1}{2}(\nabla^{\lambda}\nabla^{\rho}\phi)(\nabla_{\lambda}\nabla_{\rho}\phi)+\frac{1}{2}(\Box\phi)^{2}-(\nabla_{\lambda}\phi\nabla_{\rho}\phi)R^{\lambda\rho}\right), \nonumber \\
{\cal E}_{\phi}&=&\nabla_{\mu}\left[\left(\alpha g^{\mu\nu}-\gamma G^{\mu\nu}\right)\nabla_{\nu}\phi\right]=\nabla_{\mu} J^{\mu}_{\phi}\,,\label{L11}\\
{\cal F}_{\phi}&=&-\frac{\gamma}{4} \left[\big(\nabla_{\mu}\nabla_{\nu}\phi n^{\mu}n^{\nu}-(\nabla^{2}\phi)\big)K+(\nabla_{\mu}\nabla_{\nu}\phi)K^{\mu\nu}\right].\label{L12}
\end{eqnarray}

\section{Explicit expressions for the equations of motion}\label{AB}

The following expressions are obtained from the equations (\ref{eom1})-(\ref{eom2}):

\begin{eqnarray}
A_{t}'+\left(m^2-\frac{4\,J\,r^4\,\rho^2}{L^{4}}\right)\,p &=& 0,\label{L18} \\
\frac{\rho ''}{L^{2}} +\left(\frac{f'}{f}+\frac{1}{r}\right)\,\frac{\rho'}{L^{2}} - \left(\frac{4\,J\,r^2\,p^2}{fL^{4}}+\frac{m^2}{r^2\,f}\right)\,\rho-\frac{B}{r^2\,f} &=& 0,\label{L19} \\
A_t''-\frac{A_t'}{r}+\frac{\lambda^2}{4}\,\left(p'-\frac{p}{r}\right)&=& 0.\label{L20}
\end{eqnarray}

\section{Tensor perturbation to bulk and shear viscosity}\label{viscosity}

Here, the transport coefficients $\zeta$ and $\eta$ will be presented. To calculate these coefficients, we will perform tensor perturbations, which will be carried out inspired by the scenarios of \cite{DeWolfe:1999cp,Csaki:2000fc,Gasperini:2007zz,Santos:2020egn,Chakrabarti:2010xy,Brito:2018pwe,Brito:2019ose,Fu:2019xtx} for bulk and shear viscosity in the Horndeski equation (\ref{LEH}):

\subsection{Bulk viscosity entropy density ratio with magnetic field}

For the bulk viscosity with the metric 
\begin{eqnarray}
&&ds^{2}=h_{00}[z,t]dt^{2}+h_{11}[z,t]\left(dx^{2}+dy^{2}+dw^{2}\right)+h_{22}[z,t]dz^{2},\label{pertur}\\
&&h_{00}[z,t]=-\frac{L^{2}f(z)}{z^{2}}\Pi(z,t),\quad h_{11}[z,t]=\frac{L^{2}}{z^{2}}\chi(z,t),\quad h_{22}[z,t]=\frac{L^{2}}{z^{2}f(z)}\Gamma(z,t)\label{PT3},
\end{eqnarray}
and following the steps presented by Refs. \cite{DeWolfe:1999cp,Csaki:2000fc,Gasperini:2007zz,Santos:2020egn,Chakrabarti:2010xy,Brito:2018pwe,Brito:2019ose,Fu:2019xtx}, considering the first-order perturbations $\delta^{(1)}g_{\mu\nu}=h_{\mu\nu}$ with $h_{\mu\nu}$, where
\begin{eqnarray}
\delta^{(1)}R_{ij}&=&\partial_{\mu}(\delta^{(1)}\Gamma^{\mu}_{ij})-\partial_{i}(\delta^{(1)}\Gamma^{\mu}_{j\mu})+(\delta^{(1)}\Gamma^{\mu}_{\mu\rho})\Gamma^{\rho}_{ij}\nonumber\\
&&+\Gamma^{\mu}_{\mu\rho}(\delta^{(1)}\Gamma^{\rho}_{ij})-(\delta^{(1)}\Gamma^{\mu}_{i\rho})\Gamma^{\rho}_{\mu j}-\Gamma^{\mu}_{i\rho}(\delta^{(1)}\Gamma^{\rho}_{\mu j}),\\
\delta^{(1)}\Gamma^{k}_{ij}&=&\frac{1}{2}(\partial_{i}h^{k}_{j}+\partial_{j}h^{k}_{i}-\partial^{k}h_{ij}),
\end{eqnarray}
we can write the transverse and traceless (TT) tensor perturbation in a general way to the bulk viscosity in Horndeski gravity, where $\partial_{\alpha}h_{\mu\nu}=0$ and $h\equiv\eta^{\mu\nu}h_{\mu\nu}=0$. After an algebraic combination of the equations ${\cal E}_{tz}$ with ${\cal E}_{xx}={\cal E}_{yy}={\cal E}_{ww}$ and taking the terms in $Tr(\chi\ddot{\chi})$, $Tr(\chi^{'}\ddot{\chi})$ and $Tr(\chi^{''}\ddot{\chi})$ (for more details about this see \cite{Gasperini:2007zz}), we have
\begin{eqnarray}
\alpha^{2}L^{4}z(\alpha-\gamma \Lambda)f^{2}(z)(3\chi^{'}(z)-z \chi^{''}(z))+12(\alpha+\gamma\Lambda)\gamma^{2}z^{2}f^{2}(z)\ddot{\chi}(z)=0,
\end{eqnarray}
where now $(')$ (resp. $(\dot{})$) denotes the derivative respect to $z$ (resp. $t$). Using the ansatz: 
\begin{eqnarray}
&&\chi(z,t)=e^{-i\omega t}\varphi(z),\\
&&\varphi(z)=\exp\left(-i\omega J\ln\left(\frac{144\gamma^{2}z^{4}f(z)}{{\sqrt{3}\cal G}}\right)\right),
\end{eqnarray}
owe obtain
\begin{eqnarray}
J=\frac{1}{2\pi T}\sqrt{\frac{\alpha+\gamma\Lambda}{\alpha-\gamma\Lambda}}.
\end{eqnarray}
At this point we must evaluate the Lagrangian (\ref{eq:Lhorn}) using the metric (\ref{pertur}) and expand it up to a quadratic term, yielding the expression:
\begin{eqnarray}
&&{\cal H}_{bulk}=-96\alpha\gamma L^{2}(5\alpha+3\gamma\Lambda)f(z)(1+\chi(z,t))^{2}+432\alpha\gamma^{2}z^{2}(1+\chi(z,t))\ddot{\chi}(z,t)\cr
&&+288\alpha^{2}\gamma L^{2}f(z)(1+\chi(z,t))\chi^{'}(z,t)-24\alpha\gamma zf(z)(1+\chi(z,t))\chi^{''}(z,t)\cr
&&+216\gamma^{2}z^{2}(\alpha+\gamma\Lambda)(1+\chi(z,t))\ddot{\chi}(z,t)+648\gamma^{2}zf^{2}(z)(\alpha+\gamma\Lambda)(1+\chi(z,t))\chi^{'}(z,t)\cr
&&-18\gamma^{2}z^{2}f^{2}(z)(\alpha+\gamma\Lambda)\chi^{'2}(z,t),
\end{eqnarray}
and collecting the quadratic terms, we have
\begin{eqnarray}
&&{\cal H}_{bulk}=-{\cal M}_{1}\chi^{2}(z,t)+{\cal M}_{2}\chi(z,t)\ddot{\chi}(z,t)+{\cal M}_{3}\chi(z,t)\ddot{\chi}(z,t)+{\cal M}_{4}\chi(z,t)\chi^{'}(z,t)-\cr
&&{ \cal M}_{5}\chi(z,t)\chi^{''}(z,t)-{\cal M}_{6}\chi^{'2}(z,t)+{\cal M}_{7}\chi(z,t)\chi^{'}(z,t),\label{eq:Hbulk-zeta}
\end{eqnarray}
with
\begin{eqnarray}
&&{\cal M}_{1}=\frac{9\gamma^{2}}{8\alpha z^{3}}(5\alpha+3\gamma\Lambda)f(z),\quad {\cal M}_{2}=\frac{9\gamma^{3}}{\alpha L^{2}z},\quad
{\cal M}_{3}=\frac{9\gamma^{3}}{4\alpha^{2}L^{2}z}(\alpha+\gamma\Lambda),\nonumber\\
&&{\cal M}_{4}=\frac{6\gamma^{2}f(z)}{z^{3}},\quad
{\cal M}_{5}=\frac{9\gamma^{2} f(z)}{\alpha L^{2}z^{2}},\quad
{\cal M}_{6}=\frac{3\gamma^{3}f^{2}(z)(\alpha+\gamma\Lambda)}{16\alpha^{2}L^{2}z},\nonumber\\
&&{\cal M}_{7}=\frac{27\gamma^{3}f^{2}(z)(\alpha+\gamma\Lambda)}{4\alpha^{2}L^{2}z}.\label{eq:quad-zeta}
\end{eqnarray}
The bulk viscosity is determined from the term ${\cal M}_{4}\chi(z,t)\chi^{'}(z,t)$ from eqs. (\ref{eq:Hbulk-zeta})-(\ref{eq:quad-zeta}), given by
\begin{eqnarray}
\zeta=\frac{\sqrt{3}}{24\pi}\frac{{\cal G}}{4z^{3}_{h}}\sqrt{\frac{\alpha+\gamma\Lambda}{\alpha-\gamma\Lambda}},
\end{eqnarray}
with
\begin{eqnarray}\label{G}
{\cal G}=\frac{4z^{3}_{h}S}{\mathcal{F}},
\end{eqnarray}
where
\begin{eqnarray}\label{F}
\mathcal{F}&=&1+\frac{1}{T}\left(\frac{B^{2}\cos^{2}(\theta{'})b(\theta{'})}{5m^{2}\rho^{2}}(4\pi T)^{3}-q(\theta^{'})\left(\frac{\pi T}{3}\right)\right)\nonumber\\
&-&\frac{\sec(\theta{'})}{\left(1-\frac{\xi}{4}\right)T}\left(-\frac{B^{2}\cos^{2}(\theta{'})b(\theta{'})}{2m^{2}\rho^{2}}(\pi T)^{2}-\frac{q(\theta^{'})}{2}\right).
\end{eqnarray}
Finally, through an algebraic manipulation, we have that
\begin{eqnarray}
&&\frac{\zeta}{S}=\frac{\sqrt{3}}{24\pi\mathcal{F}}\sqrt{\frac{\alpha+\gamma\Lambda}{\alpha-\gamma\Lambda}}.
\end{eqnarray}

\subsection{Shear viscosity and entropy density ratio with magnetic field}
For the $\eta/S$ ratio, we consider the tensor perturbation in the  $xy$ metric component fluctuates. The holographic dictionary maps off-diagonal fluctuations in the bulk metric onto off-diagonal components of the dual energy-momentum tensor. In the linear regime, these fluctuations are associated with shear waves in the boundary fluid, which have a corresponding metric:
\begin{eqnarray}
ds^{2}=\frac{L^{2}}{z^{2}}\left(-f(z)dt^{2}+dw^{2}+dx^{2}+dy^{2}+2\Psi(z,t)dxdy+\frac{dz^{2}}{f(z)}\right).\label{pertur1}
\end{eqnarray}
Substituting this metric in the Horndeski equation (${\cal E}_{\mu\nu}=0$) for $\mu=x$ and $\nu=y$, one obtains:
\begin{eqnarray}
&&{\cal P}_{1}\Psi^{''}(z,t)+{\cal P}_{2}\Psi^{'}(z,t)+{\cal P}_{3}\ddot{\Psi}(z,t)=0\,,
\end{eqnarray}
where we defined
\begin{eqnarray}
&&{\cal P}_{1}=36\gamma^{2}(\alpha-\gamma\Lambda)f^{2}(z),\quad {\cal P}_{2}=-\gamma(\alpha-\gamma\Lambda)f(z)(3\alpha L^{2}-6\gamma z^{4}/z^{4}_{h}),\cr
&&{\cal P}_{3}=-36\gamma^{2}z(3\alpha+\gamma\Lambda).
\end{eqnarray}
Via the ansatz:
\begin{eqnarray}
&&\Psi(z,t)=e^{-i\omega t}\Phi(z),\\
&&\Phi(z)=\exp\left(-i\omega K\ln\left(\frac{6\gamma^{2}z^{4}f(z)}{{\cal G}}\right)\right),
\end{eqnarray}
we have that
\begin{eqnarray}
K=\frac{1}{4\pi T}\sqrt{\frac{3\alpha+\gamma\Lambda}{\alpha-\gamma\Lambda}}.
\end{eqnarray}
At this point, we must evaluate the Lagrangian (\ref{eq:Lhorn}) using the metric (\ref{pertur1}) and expand it up to quadratic terms, which reads
\begin{eqnarray}
&&{\cal H}_{shear}=P_{1}\Psi^{2}(z,t)+P_{2}\dot{\Psi}(z,t)+P_{3}\Psi^{'2}(z,t)+P_{4}\Psi(z,t)\Psi^{'}(z,t),\label{eq:Hshear}
\end{eqnarray}
where
\begin{eqnarray}
&&P_{1}=\frac{2\gamma^{2}}{z^{5}}\left(-486 \alpha\gamma\frac{z^4}{z^{4}_{h}}-\alpha  L^2(\alpha -48 \gamma\Lambda)\right), \quad P_{2}=-\frac{108\gamma^{2}(3 \alpha+\gamma\Lambda)}{z^{2}f(z) (7 \alpha+\gamma\Lambda)},\nonumber\\
&&P_{3}=\frac{6\gamma^{2}}{z^{3}f(z)},\quad P_{4}=\frac{24\gamma^{2}(\alpha+\gamma\Lambda)}{z^{4}(7 \alpha+\gamma\Lambda)}.\label{eq:quad-shear}
\end{eqnarray}
The viscosity is determined from the term $P_{3}\Psi(z,t)\Psi^{'}(z,t)$ from eqs. (\ref{eq:Hshear})-(\ref{eq:quad-shear}), given by
\begin{eqnarray}
\eta=\frac{1}{4\pi}\frac{{\cal G}}{4z^{3}_{h}}\sqrt{\frac{3\alpha+\gamma\Lambda}{\alpha-\gamma\Lambda}},
\end{eqnarray}
where for ${\cal{G}}$ and $\mathcal{F}$ given in (\ref{G}) and (\ref{F}) previously, we can provide:
\begin{eqnarray}
\frac{\eta}{S}=\frac{1}{4\pi\mathcal{F}}\sqrt{\frac{3\alpha+\gamma\Lambda}{\alpha-\gamma\Lambda}}.
\end{eqnarray}



\begin{thebibliography}{99}
  \bibitem{Maldacena:1997re} 
  J.~M.~Maldacena,
  {\it The Large N limit of superconformal field theories and supergravity},
  Int.\ J.\ Theor.\ Phys.\  {\bf 38}, 1113 (1999)
  [Adv.\ Theor.\ Math.\ Phys.\  {\bf 2}, 231 (1998)]
    [hep-th/9711200].
  
\bibitem{Witten:1998qj}
E.~Witten,
Adv. Theor. Math. Phys. \textbf{2} (1998), 253-291
doi:10.4310/ATMP.1998.v2.n2.a2
[arXiv:hep-th/9802150 [hep-th]].

\bibitem{Ahn:2022azl}
Y.~Ahn, M.~Baggioli, K.~B.~Huh, H.~S.~Jeong, K.~Y.~Kim and Y.~W.~Sun,
{\it Holography and magnetohydrodynamics with dynamical gauge fields},
[arXiv:2211.01760 [hep-th]].

\bibitem{Ballon-Bayona:2022uyy}
A.~Ballon-Bayona, J.~P.~Shock and D.~Zoakos,
{\it Magnetising the $ \mathcal{N} $ = 4 Super Yang-Mills plasma},
JHEP \textbf{06}, 154 (2022)
doi:10.1007/JHEP06(2022)154
[arXiv:2203.00050 [hep-th]].

\bibitem{Skokov:2009qp}
V.~Skokov, A.~Y.~Illarionov and V.~Toneev,
{\it Estimate of the magnetic field strength in heavy-ion collisions},
Int. J. Mod. Phys. A \textbf{24}, 5925-5932 (2009)
doi:10.1142/S0217751X09047570
[arXiv:0907.1396 [nucl-th]].

\bibitem{Fukushima:2016xgg}
K.~Fukushima,
{\it Evolution to the quark\textendash{}gluon plasma},
Rept. Prog. Phys. \textbf{80}, no.2, 022301 (2017)
doi:10.1088/1361-6633/80/2/022301
[arXiv:1603.02340 [nucl-th]].

\bibitem{Gubser:2008px}
S.~S.~Gubser,
{\it Breaking an Abelian gauge symmetry near a black hole horizon},
Phys. Rev. D \textbf{78}, 065034 (2008)
doi:10.1103/PhysRevD.78.065034
[arXiv:0801.2977 [hep-th]].


\bibitem{Gubser:1996de}
S.~S.~Gubser, I.~R.~Klebanov and A.~W.~Peet,
Phys. Rev. D \textbf{54} (1996), 3915-3919
doi:10.1103/PhysRevD.54.3915
[arXiv:hep-th/9602135 [hep-th]].


\bibitem{Burgess:1999vb}
C.~P.~Burgess, N.~R.~Constable and R.~C.~Myers,
JHEP \textbf{08} (1999), 017
doi:10.1088/1126-6708/1999/08/017
[arXiv:hep-th/9907188 [hep-th]].


\bibitem{Policastro:2001yc}
G.~Policastro, D.~T.~Son and A.~O.~Starinets,
Phys. Rev. Lett. \textbf{87} (2001), 081601
doi:10.1103/PhysRevLett.87.081601
[arXiv:hep-th/0104066 [hep-th]].


\bibitem{Policastro:2002se}
G.~Policastro, D.~T.~Son and A.~O.~Starinets,
JHEP \textbf{09} (2002), 043
doi:10.1088/1126-6708/2002/09/043
[arXiv:hep-th/0205052 [hep-th]].








\bibitem{Takayanagi:2011zk} 
T.~Takayanagi,
``Holographic Dual of BCFT,''
Phys.\ Rev.\ Lett.\  {\bf 107}, 101602 (2011),
[arXiv:1105.5165 [hep-th]].

\bibitem{Fujita:2011fp} 
M.~Fujita, T.~Takayanagi and E.~Tonni,
``Aspects of AdS/BCFT,''
JHEP {\bf 1111}, 043 (2011),
[arXiv:1108.5152 [hep-th]].

\bibitem{Fujita:2012fp}
M.~Fujita, M.~Kaminski and A.~Karch,
{\it SL(2,Z) Action on AdS/BCFT and Hall Conductivities},
JHEP \textbf{07}, 150 (2012)
doi:10.1007/JHEP07(2012)150
[arXiv:1204.0012 [hep-th]].

\bibitem{Melnikov:2012tb}
D.~Melnikov, E.~Orazi and P.~Sodano,
{\it On the AdS/BCFT Approach to Quantum Hall Systems},
JHEP \textbf{05}, 116 (2013)
doi:10.1007/JHEP05(2013)116
[arXiv:1211.1416 [hep-th]].

\bibitem{Miao:2018qkc}{R.~X.~Miao, 
{\it Holographic BCFT with Dirichlet Boundary Condition.} 
JHEP {\bf 1902}, 025 (2019), [arXiv:1806.10777 [hep-th]].}
 
\bibitem{Magan:2014dwa}
J.~M.~Mag\'an, D.~Melnikov and M.~R.~O.~Silva,
{\it Black Holes in AdS/BCFT and Fluid/Gravity Correspondence},
JHEP \textbf{11} (2014), 069
doi:10.1007/JHEP11(2014)069
[arXiv:1408.2580 [hep-th]].

\bibitem{dosSantos:2022scy}
F.~F.~dos Santos,
{\it AdS/BCFT correspondence and BTZ black hole within electric field},
JHAP \textbf{4}, no.1, 81-92 (2022)
doi:10.22128/jhap.2022.504.1018
[arXiv:2206.09502 [hep-th]].

\bibitem{Sokoliuk:2022llp}
O.~Sokoliuk, F.~F.~Santos and A.~Baransky,
{\it AdS/BCFT correspondence and Lovelock theory in the presence of canonical scalar field},
[arXiv:2206.04054 [hep-th]].

\bibitem{Santos:2021orr}
F.~F.~Santos, E.~F.~Capossoli and H.~Boschi-Filho,
``AdS/BCFT correspondence and BTZ black hole thermodynamics within Horndeski gravity,''
Phys. Rev. D \textbf{104}, no.6, 066014 (2021)
[arXiv:2105.03802 [hep-th]].

\bibitem{Santos:2023flb}
F.~F.~Santos, M.~Bravo-Gaete, O.~Sokoliuk and A.~Baransky,
Fortsch. Phys. \textbf{71}, no.12, 2300008 (2023)
doi:10.1002/prop.202300008
[arXiv:2301.03121 [hep-th]].

\bibitem{Jeong:2023hrb}
H.~S.~Jeong, K.~Y.~Kim and Y.~W.~Sun,
[arXiv:2305.18122 [hep-th]].

\bibitem{Ryu:2006bv}
S.~Ryu and T.~Takayanagi,
{\it Holographic derivation of entanglement entropy from AdS/CFT},
Phys. Rev. Lett. \textbf{96}, 181602 (2006)
doi:10.1103/PhysRevLett.96.181602
[arXiv:hep-th/0603001 [hep-th]].

\bibitem{DosSantos:2022exb}
F.~F.~Dos Santos,
{\it Entanglement entropy in Horndeski gravity},
JHAP \textbf{3}, no.1, 1-14 (2022)
doi:10.22128/jhap.2022.488.1015
[arXiv:2201.02500 [hep-th]].

\bibitem{Santos:2023eqp}
F.~F.~Santos, B.~Pourhassan and E.~Saridakis,
{\it de Sitter versus anti-de Sitter in Horndeski-like gravity},
[arXiv:2305.05794 [hep-th]].

\bibitem{Randall:1999vf}
L.~Randall and R.~Sundrum,
{\it An Alternative to compactification},
Phys. Rev. Lett. \textbf{83}, 4690-4693 (1999)
doi:10.1103/PhysRevLett.83.4690
[arXiv:hep-th/9906064 [hep-th]].

\bibitem{Horndeski:1974wa}
  G.~W.~Horndeski,
  Int.\ J.\ Theor.\ Phys.\  {\bf 10}, 363 (1974).



\bibitem{Deffayet:2009mn}
C.~Deffayet, S.~Deser and G.~Esposito-Farese,
Phys. Rev. D \textbf{80} (2009), 064015
doi:10.1103/PhysRevD.80.064015
[arXiv:0906.1967 [gr-qc]].

\bibitem{Charmousis:2014mia}
C.~Charmousis,
Lect. Notes Phys. \textbf{892} (2015), 25-56
doi:10.1007/978-3-319-10070-8\_2
[arXiv:1405.1612 [gr-qc]].

\bibitem{Rinaldi:2012vy}
M.~Rinaldi,
{\it Black holes with non-minimal derivative coupling},
Phys. Rev. D \textbf{86} (2012), 084048
doi:10.1103/PhysRevD.86.084048
[arXiv:1208.0103 [gr-qc]].


\bibitem{Feng:2015oea}
X.~H.~Feng, H.~S.~Liu, H.~L\"u and C.~N.~Pope,
{\it Black Hole Entropy and Viscosity Bound in Horndeski Gravity},
JHEP \textbf{11}, 176 (2015)
doi:10.1007/JHEP11(2015)176
[arXiv:1509.07142 [hep-th]].

\bibitem{Bravo-Gaete:2022lno}
M.~Bravo-Gaete, F.~F.~Santos and H.~Boschi-Filho,
{\it Shear viscosity from black holes in generalized scalar-tensor theories in arbitrary dimensions},
Phys. Rev. D \textbf{106}, no.6, 066010 (2022)
doi:10.1103/PhysRevD.106.066010
[arXiv:2201.07961 [hep-th]].

\bibitem{Bravo-Gaete:2023iry}
M.~Bravo-Gaete, L.~Guajardo and F.~F.~Santos,
{\it Exploring the shear viscosity in four-dimensional planar black holes beyond general relativity},
Phys. Rev. D \textbf{107}, no.10, 104032 (2023)
doi:10.1103/PhysRevD.107.104032
[arXiv:2303.07493 [hep-th]].

\bibitem{Bravo-Gaete:2020lzs}
M.~Bravo-Gaete and F.~F.~Santos,
{\it Complexity of four-dimensional hairy anti-de-Sitter black holes with a rotating string and shear viscosity in generalized scalar\textendash{}tensor theories},
Eur. Phys. J. C \textbf{82}, no.2, 101 (2022)
doi:10.1140/epjc/s10052-022-10064-y
[arXiv:2010.10942 [hep-th]].

\bibitem{Parnachev:2005hh}
A.~Parnachev and A.~Starinets,
{\it The Silence of the little strings},
JHEP \textbf{10}, 027 (2005)
doi:10.1088/1126-6708/2005/10/027
[arXiv:hep-th/0506144 [hep-th]].

\bibitem{Benincasa:2005iv}
P.~Benincasa, A.~Buchel and A.~O.~Starinets,
{\it Sound waves in strongly coupled non-conformal gauge theory plasma},
Nucl. Phys. B \textbf{733}, 160-187 (2006)
doi:10.1016/j.nuclphysb.2005.11.005
[arXiv:hep-th/0507026 [hep-th]].

\bibitem{Buchel:2005cv}
A.~Buchel,
{\it Transport properties of cascading gauge theories},
Phys. Rev. D \textbf{72}, 106002 (2005)
doi:10.1103/PhysRevD.72.106002
[arXiv:hep-th/0509083 [hep-th]].

\bibitem{Benincasa:2006ei}
P.~Benincasa and A.~Buchel,
{\it Hydrodynamics of Sakai-Sugimoto model in the quenched approximation},
Phys. Lett. B \textbf{640}, 108-115 (2006)
doi:10.1016/j.physletb.2006.07.043
[arXiv:hep-th/0605076 [hep-th]].

\bibitem{Mas:2007ng}
J.~Mas and J.~Tarrio,
{\it Hydrodynamics from the Dp-brane},
JHEP \textbf{05}, 036 (2007)
doi:10.1088/1126-6708/2007/05/036
[arXiv:hep-th/0703093 [hep-th]].



\bibitem{Bravo-Gaete:2021hlc}
M.~Bravo-Gaete and M.~M.~Stetsko,
Phys. Rev. D \textbf{105}, no.2, 024038 (2022)
doi:10.1103/PhysRevD.105.024038
[arXiv:2111.10925 [hep-th]].

\bibitem{Torres-Rincon:2012sda}
J.~M.~Torres-Rincon,
{\it Hadronic transport coefficients from effective field theories},
doi:10.1007/978-3-319-00425-9
[arXiv:1205.0782 [hep-ph]].

\bibitem{Mishra:2020onx}
A.~K.~Mishra,
{\it Exploring the self interacting dark matter properties from low redshift observations},
Eur. Phys. J. C \textbf{82}, no.11, 1060 (2022)
doi:10.1140/epjc/s10052-022-10907-8
[arXiv:2002.11652 [astro-ph.CO]].



\bibitem{Gubser:2008sz}
S.~S.~Gubser, S.~S.~Pufu and F.~D.~Rocha,
{\it Bulk viscosity of strongly coupled plasmas with holographic duals},
JHEP \textbf{08}, 085 (2008)
doi:10.1088/1126-6708/2008/08/085
[arXiv:0806.0407 [hep-th]].

\bibitem{Rebhan:2011vd}
A.~Rebhan and D.~Steineder,
{\it Violation of the Holographic Viscosity Bound in a Strongly Coupled Anisotropic Plasma},
Phys. Rev. Lett. \textbf{108}, 021601 (2012)
doi:10.1103/PhysRevLett.108.021601
[arXiv:1110.6825 [hep-th]].



\bibitem{Bluhm:2007nu}
M.~Bluhm, B.~Kampfer, R.~Schulze, D.~Seipt and U.~Heinz,
Phys. Rev. C \textbf{76}, 034901 (2007)
doi:10.1103/PhysRevC.76.034901
[arXiv:0705.0397 [hep-ph]].

\bibitem{Cherman:2009tw}
A.~Cherman, T.~D.~Cohen and A.~Nellore,
Phys. Rev. D \textbf{80}, 066003 (2009)
doi:10.1103/PhysRevD.80.066003
[arXiv:0905.0903 [hep-th]].





\bibitem{Iida:2022hyy}
K.~Iida and E.~Itou,
PTEP \textbf{2022}, no.11, 111B01 (2022)
doi:10.1093/ptep/ptac137
[arXiv:2207.01253 [hep-ph]].


\bibitem{Itou:2023pcl}
E.~Itou and K.~Iida,
[arXiv:2311.15259 [hep-lat]].

\bibitem{Brandt:2022hwy}
B.~B.~Brandt, F.~Cuteri and G.~Endrodi,
JHEP \textbf{07}, 055 (2023)
doi:10.1007/JHEP07(2023)055
[arXiv:2212.14016 [hep-lat]].

\bibitem{Abbott:2023coj}
R.~Abbott \textit{et al.} [NPLQCD],
Phys. Rev. D \textbf{108}, no.11, 114506 (2023)
doi:10.1103/PhysRevD.108.114506
[arXiv:2307.15014 [hep-lat]].

\bibitem{Buchel:2007mf}
A.~Buchel,
{\it Bulk viscosity of gauge theory plasma at strong coupling},
Phys. Lett. B \textbf{663}, 286-289 (2008)
doi:10.1016/j.physletb.2008.03.069
[arXiv:0708.3459 [hep-th]].

\bibitem{Kharzeev:2007wb}
D.~Kharzeev and K.~Tuchin,
{\it Bulk viscosity of QCD matter near the critical temperature},
JHEP \textbf{09}, 093 (2008)
doi:10.1088/1126-6708/2008/09/093
[arXiv:0705.4280 [hep-ph]].

\bibitem{Karsch:2007jc}
F.~Karsch, D.~Kharzeev and K.~Tuchin,
{\it Universal properties of bulk viscosity near the QCD phase transition},
Phys. Lett. B \textbf{663}, 217-221 (2008)
doi:10.1016/j.physletb.2008.01.080
[arXiv:0711.0914 [hep-ph]].

\bibitem{Gubser:2008yx}
S.~S.~Gubser, A.~Nellore, S.~S.~Pufu and F.~D.~Rocha,
{\it Thermodynamics and bulk viscosity of approximate black hole duals to finite temperature quantum chromodynamics},
Phys. Rev. Lett. \textbf{101}, 131601 (2008)
doi:10.1103/PhysRevLett.101.131601
[arXiv:0804.1950 [hep-th]].


\bibitem{Kharzeev:2007jp}
D.~E.~Kharzeev, L.~D.~McLerran and H.~J.~Warringa,
{\it The Effects of topological charge change in heavy ion collisions: 'Event by event P and CP violation'},
Nucl. Phys. A \textbf{803} (2008), 227-253
doi:10.1016/j.nuclphysa.2008.02.298
[arXiv:0711.0950 [hep-ph]].

\bibitem{Fukushima:2008xe}
K.~Fukushima, D.~E.~Kharzeev and H.~J.~Warringa,
{\it The Chiral Magnetic Effect},
Phys. Rev. D \textbf{78} (2008), 074033
doi:10.1103/PhysRevD.78.074033
[arXiv:0808.3382 [hep-ph]].


\bibitem{Endrodi:2018ikq}
G.~Endrodi, M.~Kaminski, A.~Schafer, J.~Wu and L.~Yaffe,
{\it Universal Magnetoresponse in QCD and $\mathcal{N}=4$ SYM},
JHEP \textbf{09}, 070 (2018)
doi:10.1007/JHEP09(2018)070
[arXiv:1806.09632 [hep-th]].





\bibitem{Babichev:2013cya}
E.~Babichev and C.~Charmousis,
{\it Dressing a black hole with a time-dependent Galileon},
JHEP \textbf{08} (2014), 106
doi:10.1007/JHEP08(2014)106
[arXiv:1312.3204 [gr-qc]].

\bibitem{Anabalon:2013oea}
A.~Anabalon, A.~Cisterna and J.~Oliva,
{\it Asymptotically locally AdS and flat black holes in Horndeski theory},
Phys. Rev. D \textbf{89} (2014), 084050
doi:10.1103/PhysRevD.89.084050
[arXiv:1312.3597 [gr-qc]].


\bibitem{Bravo-Gaete:2014haa}
M.~Bravo-Gaete and M.~Hassaine,
{\it Thermodynamics of a BTZ black hole solution with an Horndeski source},
Phys. Rev. D \textbf{90} (2014) no.2, 024008
doi:10.1103/PhysRevD.90.024008
[arXiv:1405.4935 [hep-th]].

\bibitem{Bravo-Gaete:2013dca}
M.~Bravo-Gaete and M.~Hassaine,
{\it Lifshitz black holes with a time-dependent scalar field in a Horndeski theory},
Phys. Rev. D \textbf{89} (2014), 104028
doi:10.1103/PhysRevD.89.104028
[arXiv:1312.7736 [hep-th]].




\bibitem{Brihaye:2016lin}
Y.~Brihaye, A.~Cisterna and C.~Erices,
{\it Boson stars in biscalar extensions of Horndeski gravity},
Phys. Rev. D \textbf{93} (2016) no.12, 124057
doi:10.1103/PhysRevD.93.124057
[arXiv:1604.02121 [hep-th]].

\bibitem{Cisterna:2015yla}
A.~Cisterna, T.~Delsate and M.~Rinaldi,
{\it Neutron stars in general second order scalar-tensor theory: The case of nonminimal derivative coupling},
Phys. Rev. D \textbf{92} (2015) no.4, 044050
doi:10.1103/PhysRevD.92.044050
[arXiv:1504.05189 [gr-qc]].

\bibitem{Cisterna:2016vdx}
A.~Cisterna, T.~Delsate, L.~Ducobu and M.~Rinaldi,
{\it Slowly rotating neutron stars in the nonminimal derivative coupling sector of Horndeski gravity},
Phys. Rev. D \textbf{93} (2016) no.8, 084046
doi:10.1103/PhysRevD.93.084046
[arXiv:1602.06939 [gr-qc]].

\bibitem{Arratia:2020hoy}
E.~Arratia, C.~Corral, J.~Figueroa and L.~Sanhueza,
{\it Hairy Taub-NUT/bolt-AdS solutions in Horndeski theory},
Phys. Rev. D \textbf{103} (2021) no.6, 064068
doi:10.1103/PhysRevD.103.064068
[arXiv:2010.02460 [hep-th]].

\bibitem{Caceres:2023gfa}
N.~C\'aceres, C.~Corral, F.~Diaz and R.~Olea,
[arXiv:2311.04054 [hep-th]].

\bibitem{Feng:2018sqm}
X.~H.~Feng and H.~S.~Liu,
{\it Holographic Complexity Growth Rate in Horndeski Theory},
Eur. Phys. J. C \textbf{79} (2019) no.1, 40
doi:10.1140/epjc/s10052-019-6547-4
[arXiv:1811.03303 [hep-th]].

\bibitem{Hui:2012qt}
L.~Hui and A.~Nicolis,
{\it No-Hair Theorem for the Galileon},
Phys. Rev. Lett. \textbf{110} (2013), 241104
doi:10.1103/PhysRevLett.110.241104
[arXiv:1202.1296 [hep-th]].

\bibitem{Brito:2019ose}
F.~Brito and F.~Santos,
``Black branes in asymptotically Lifshitz spacetime and viscosity/entropy ratios in Horndeski gravity,''
EPL \textbf{129}, no.5, 50003 (2020),
[arXiv:1901.06770 [hep-th]].




\bibitem{Ghotbabadi:2021mus}
B.~B.~Ghotbabadi, A.~Sheykhi, G.~H.~Bordbar and A.~Montakhab,
{\it Holographic paramagnetic-ferromagnetic phase transition of Power-Maxwell-Gauss-Bonnet black holes},
[arXiv:2102.08053 [hep-th]].


\bibitem{Cai:2014oca}
R.~G.~Cai and R.~Q.~Yang,
{\it Paramagnetism-Ferromagnetism Phase Transition in a Dyonic Black Hole}
Phys. Rev. D \textbf{90}, no.8, 081901 (2014)
[arXiv:1404.2856 [hep-th]].

\bibitem{Cai:2015bsa}
R.~G.~Cai and R.~Q.~Yang,
{\it Antisymmetric tensor field and spontaneous magnetization in holographic duality},
Phys. Rev. D \textbf{92} (2015) no.4, 046001
[arXiv:1504.00855 [hep-th]].




\bibitem{Hartnoll:2008kx}
S.~A.~Hartnoll, C.~P.~Herzog and G.~T.~Horowitz,
{\it Holographic Superconductors},
JHEP \textbf{12}, 015 (2008)
doi:10.1088/1126-6708/2008/12/015
[arXiv:0810.1563 [hep-th]].

\bibitem{Hartnoll:2009sz}
S.~A.~Hartnoll,
{\it Lectures on holographic methods for condensed matter physics},
Class. Quant. Grav. \textbf{26}, 224002 (2009)
doi:10.1088/0264-9381/26/22/224002
[arXiv:0903.3246 [hep-th]].






\bibitem{Karsch}
F.~Karsch and E.~Laermann, in Quark Gluon Plasma 3, edited by R.C.Hwa and X.N.Wang (World Scientific, Singapore,2004),p.1.

\bibitem{ALICE:2019sgg}
S.~Acharya \textit{et al.} [ALICE],
{\it Probing the effects of strong electromagnetic fields with charge-dependent directed flow in Pb-Pb collisions at the LHC},
Phys. Rev. Lett. \textbf{125} (2020) no.2, 022301
doi:10.1103/PhysRevLett.125.022301
[arXiv:1910.14406 [nucl-ex]].


\bibitem{STAR:2019clv}
J.~Adam \textit{et al.} [STAR],
{\it First Observation of the Directed Flow of $D^{0}$ and $\overline{D^0}$ in Au+Au Collisions at $\sqrt{s_{\rm NN}}$ = 200 GeV},
Phys. Rev. Lett. \textbf{123} (2019) no.16, 162301
doi:10.1103/PhysRevLett.123.162301
[arXiv:1905.02052 [nucl-ex]].





\bibitem{Bali:2014kia}
G.~S.~Bali, F.~Bruckmann, G.~Endr\"odi, S.~D.~Katz and A.~Sch\"afer,
{\it The QCD equation of state in background magnetic fields},
JHEP \textbf{08}, 177 (2014)
doi:10.1007/JHEP08(2014)177
[arXiv:1406.0269 [hep-lat]].






\bibitem{Brito:2018pwe}
F.~A.~Brito and F.~F.~Santos,
{\it Braneworlds in Horndeski gravity},
Eur. Phys. J. Plus \textbf{137}, no.9, 1051 (2022)
doi:10.1140/epjp/s13360-022-03270-w
[arXiv:1810.08196 [hep-th]].

\bibitem{DeWolfe:1999cp} 
  O.~DeWolfe, D.~Z.~Freedman, S.~S.~Gubser and A.~Karch,
  {\it Modeling the fifth-dimension with scalars and gravity},
  Phys.\ Rev.\ D {\bf 62}, 046008 (2000),
  [hep-th/9909134].
  
\bibitem{Csaki:2000fc} 
  C.~Csaki, J.~Erlich, T.~J.~Hollowood and Y.~Shirman,
  {\it Universal aspects of gravity localized on thick branes},
  Nucl.\ Phys.\ B {\bf 581}, 309 (2000),
  [hep-th/0001033].
  
\bibitem{Gasperini:2007zz}
M.~Gasperini,
{\it Elements of string cosmology}.

\bibitem{Santos:2020egn}
F.~F.~Santos,
{\it Aplica\c{c}\~oes do Setor John da Gravidade de Horndeski nos Cen\'arios de Brana Negra e Rela\c{c}\~ao de viscosidade/entropia, Mundo Brana e Cosmologia (In Portuguese)},
[arXiv:2006.06550 [hep-th]].

  \bibitem{Chakrabarti:2010xy} 
  S.~K.~Chakrabarti, S.~Chakrabortty and S.~Jain,
  {\it Proof of universality of electrical conductivity at finite chemical potential},
  JHEP {\bf 1102}, 073 (2011),
  [arXiv:1011.3499 [hep-th]].

\bibitem{Fu:2019xtx} 
  Q.~M.~Fu, H.~Yu, L.~Zhao and Y.~X.~Liu,
  {\it Thick brane in reduced Horndeski theory},
  Phys.\ Rev.\ D {\bf 100}, no. 12, 124057 (2019),
  [arXiv:1907.12049 [gr-qc]].

\bibitem{Nicolis:2008in}
A.~Nicolis, R.~Rattazzi and E.~Trincherini,
{\it The Galileon as a local modification of gravity},
Phys. Rev. D \textbf{79}, 064036 (2009)
doi:10.1103/PhysRevD.79.064036
[arXiv:0811.2197 [hep-th]].


 
\end{thebibliography}
\end{document}